\documentclass[amsmath,amssymb,aps,nofootinbib,showkeys,superscriptaddress]{revtex4-2}

\usepackage{graphicx}                   
\usepackage[lofdepth,lotdepth]{subfig}	
\usepackage{dcolumn}                    
\usepackage{bm}                         
\usepackage{booktabs}                   

\graphicspath{{figures/}}

\begin{document}
\preprint{APS/123-QED}

\title{From nearly homogeneous to core-peaking suspensions: insight in suspension pipe flows using MRI and DNS}

\author{Willian Hogendoorn}
\email{w.j.hogendoorn@tudelft.nl}
\affiliation{
Delft University of Technology, Multiphase Systems (3ME-P\&E), Leeghwaterstraat 39, 2628 CB Delft, The Netherlands\\
}

\author{Wim-Paul Breugem\footnote{Hogendoorn and Breugem contributed equally to this study.}}
\email{w.p.breugem@tudelft.nl}
\affiliation{
Delft University of Technology, Multiphase Systems (3ME-P\&E), Leeghwaterstraat 39, 2628 CB Delft, The Netherlands\\
}

\author{David Frank}
\affiliation{
University of Rostock, Institute of Fluid Mechanics, Justus-von-Liebig-Weg 2, 18059, Rostock, Germany\\
}

\author{Martin Bruschewski}
\affiliation{
University of Rostock, Institute of Fluid Mechanics, Justus-von-Liebig-Weg 2, 18059, Rostock, Germany\\
}

\author{Sven Grundmann}
\affiliation{
University of Rostock, Institute of Fluid Mechanics, Justus-von-Liebig-Weg 2, 18059, Rostock, Germany\\
}

\author{Christian Poelma}
\affiliation{
Delft University of Technology, Multiphase Systems (3ME-P\&E), Leeghwaterstraat 39, 2628 CB Delft, The Netherlands\\
}

\date{\today}

\begin{abstract}
Magnetic Resonance Imaging (MRI) experiments have been performed in conjunction with Direct Numerical Simulations (DNS) to study neutrally buoyant particle-laden pipe flows. The flows are characterized by the suspension liquid Reynolds number ($Re_s$), based on the bulk liquid velocity and suspension viscosity obtained from Eilers' correlation, the bulk solid volume fraction ($\phi_b$) and the particle-to-pipe diameter ratio ($d/D$). Six different cases have been studied, each with a unique combination of $Re_s$ and $\phi$, while $d/D$ is kept constant at 0.058. The selected cases ensure that the comparison is performed across different flow regimes, each exhibiting characteristic behavior. In general, an excellent agreement is found between experiment and simulation for the average liquid velocity and solid volume fraction profiles. Root mean square errors as low as 1.2\% and 8.4\% are found for the velocity and volume fraction profiles, respectively. This study presents, for the first time, accurate and quantitative velocity and volume fraction profiles of semi-dilute up to dense suspension flows using both experimental and numerical methods. The discrepancy between the experiments and simulations can be explained by various reasons, including a difference in particle size distribution, uncertainty in experimental parameters used as an input for the DNS, slight variations in particle roughness and frictional collisions in the simulations. Eventually, three different flow regimes are identified, based on the experimental and numerical solid volume fraction profiles. These profiles explain observations in the drag change. For \textit{low} bulk solid volume fractions a drag increase (with respect to an equal $Re_s$ single-phase case) is observed, which is linked to a nearly uniform distributed system and a layer of particles lining the pipe wall that acts as drag-enhancing rough/porous wall layer. For \textit{moderate} volume fraction distributions the drag is found to decrease, due to particle accumulation at the pipe centre. For \textit{high} volume fractions the drag is found to decrease further. For solid volume fractions of 0.4 a drag reduction higher than 25\% is found. This drag reduction is linked to the strong viscosity gradient in the radial direction, where the relatively low viscosity near the pipe wall acts as a lubrication layer between the pipe wall and the dense core.

\end{abstract}


\maketitle


\section{\label{sec:Introduction}Introduction}
Over the last decades there has been a continuously growing interest in flows of solid particles dispersed in a liquid phase \cite{guazzelli2018rheology, morris2020toward}. This interest is motivated by the fact that these suspension flows are present in a broad spectrum of processes, including food processing, manufacturing of healthcare products, and sediment transport. Understanding and modeling of suspension flow behavior in the moderate and concentrated regimes is considered to be a challenge, in particular when inertial effects can no longer be ignored. Despite significant theoretical, experimental and numerical progress, open questions still remain. The major aim of this study is twofold: first, to provide a comprehensive comparison between experimental results obtained using magnetic resonance imaging (MRI) and particle-resolved direct numerical simulations (DNS) in different suspension flow regimes; and second, to further study the physics underlying these various flow regimes using the established data set.\\

A phenomenon that complicates the flow of suspensions is shear-induced migration:
in the presence of inhomogeneous shear, such as in pressure-driven pipe flow, an initially homogeneous suspension may rearrange into a non-homogeneous mixture.
Suspended particles migrate towards the low-shear regions in the flow and form particle clusters, resulting in wall-normal concentration gradients.
In turn, these concentration gradients are responsible for strong gradients in the suspension viscosity, and thus in turn influence the shear profiles.
This behavior is observed in various experimental facilities, including (wide-gap) annular Couette systems \cite{abbott1991experimental, graham1991note, shapley2004evaluation, blanc2013microstructure, tetlow1998particle, ovarlez2006local}, rectangular channel flows \cite{hookham1986concentration, zade2018experimental, lyon1998experimental}, and pipe flow \cite{altobelli1991velocity, sinton1991nmr, kalyon1993rheological, butler1999imaging, butler1999observations}.
A connection between shear-induced migration and radial migration of particles as initially observed by \citet{segre1962behavior} was suggested by \citet{han1999particle}.
\citet{nott1994pressure} distinguish between both phenomena as the radial migration effect is inertia driven, in contrast to shear-induced migration, which is observed already in the Stokes regime.

One of the pioneering experiments reporting shear-induced migration in pipe flow with dense suspensions was performed by \citet{karnis1966kinetics} in the 1960s. 
The authors used a refractive index matched suspension to ascertain optical access. 
Their measurement system was a camera in combination with a microscope.
Velocity and concentration profiles were obtained after processing the camera images.
For increasing bulk solid volume fraction ($\phi_b>$ 0.14), the velocity profile was found to transition from a parabola to a blunted profile due to the presence of a `partial plug flow' (i.e., particles are found to accumulate at the pipe centre). 
Furthermore, the particle-to-pipe diameter ratio ($d/D$) was found to affect the migration behavior: larger particles resulted in a more pronounced velocity blunting for the same volume fraction.

This shear-induced particle migration was also observed in an experimental study in an annular Couette system by \citet{gadala1980shear}.
A consistent viscosity \textit{decrease} for higher volume fractions ($\phi_b >$ 0.3) was found, suggesting a non-homogeneous particle distribution.
The authors concluded that a concentrated suspension should be modelled using a {\em{local}} effective viscosity rather than a constant effective viscosity.

Similar behavior was observed in a rectangular channel by \citet{hookham1986concentration}.
Average velocity and concentration profiles were obtained using an adapted laser Doppler technique in combination with fluorescent particles.
Based on the data obtained, a particle accumulation at the channel centre was observed. 
In addition, the velocity profile appeared to be blunted, where the degree of blunting was found to increase with increasing volume fraction. 

The behaviour observed by \citet{hookham1986concentration} was later confirmed in the experiments by \citet{koh1994experimental}.
The authors performed laser Doppler anemometry experiments in a refractive indexed matched dense suspension in a rectangular channel.
A comparison was made with theoretical models based on shear-induced particle migration (SIM) introduced by \citet{leighton1987shear} and \citet{phillips1992constitutive}.

In the meantime, the first MRI measurements were introduced to study rheology.
Pioneering MRI measurements in particle-laden pipe flow were performed by \textcite{majors1989velocity}.
They studied a suspension with volume fractions ranging from 0.016\textendash0.10.
The focus of their study was to introduce and illustrate MRI as a quantitative and non-invasive measurement technique for suspension flows, rather than a detailed rheological study. 
Therefore, the authors refrained from a discussion about shear-induced particle migration. 

\citet{sinton1991nmr} also performed MRI measurements in both Newtonian and non-Newtonian pipe flow. They studied neutrally buoyant suspensions with volume fractions ranging from $\phi_b = 0.21$ to $0.52$ in Stokes flow. They found that the degree of blunting of the velocity profile does not only depend on the solid volume fraction, but also on the particle-to-pipe diameter ratio and the pipe length-to-diameter ratio ($L/D$), presumably related to either wall effects or an `induction length' for particle migration to reach an equilibrium particle distribution across the pipe.

In addition to these experimental studies, progress was made using theoretical modeling. For instance, a suspension balance model (SBM) was introduced by \citet{nott1994pressure}. This model was later revisited by \citet{nott2011suspension}, by adding a well-defined particle phase stress, solving the issue that the particle phase stress was identified with the particle contribution to the suspension stress. By taking this particle stress contribution explicitly into account, this model is distinct from other models, which only take into account a force acting on the particle phase. The SBM suggested that previous experimental studies used insufficient development lengths: some of the listed studies use only a few diameters of development length before characterizing the flow. The exact amount of length needed for full development of velocity {\em{and}} concentration profiles was not well established for dense suspensions. 
Therefore, further MRI measurements of neutrally buoyant suspensions at different streamwise pipe locations were performed by \citet{hampton1997migration}.
Two different $d/D$ values were studied with volume fractions ranging between $\phi_b$ = 0.10 and 0.50.
The main focus of the experiments was to investigate the required development or entrance length associated with different volume fractions.
Hence, concentration (and velocity) profiles at different pipe locations were taken.
Based on the experimental data a model was proposed to capture the streamwise concentration profile development.
Another interesting observation is that for average volume fractions of 0.20\textendash0.40 and $d/D$ = 0.0625, ordered particle layers were observed in the vicinity of the pipe wall.
The authors point out that the constraining pipe wall is likely responsible for this particle ordering.
A comparison with the SIM and SBM models was made to explain the observed flow behavior. However, neither model provided a good quantitative prediction for the obtained results.

These experiments were followed by measurements by \citet{han1999particle}.
They showed that for low volume fractions (i.e., about 6\%) inertia and particle-particle interactions should be taken into account for the modelling of the (radial) concentration profile.
They suggest that for particle Reynolds numbers ($Re_p = a u_b/\nu$, with $u_b$ the bulk mixture velocity, $a=d/2$ the particle radius, and $\nu$ the liquid phase viscosity) exceeding 0.1, inertial effects cannot be neglected for any volume fraction. 

After the pioneering work in the 1990s, very few other studies investigating suspension flow dynamics in pipe flow were reported.
More recent studies focused mainly on the effect of particles in channel or duct flow, in particular in the turbulent regime (see, e.g., the numerical studies by \citet{sharma2006turbulent, fornari2018suspensions, costa2016universal, costa2018effects}).
Also a combined experimental and numerical study on the effect of buoyant particles in turbulent duct flow was reported by \citet{zade2019buoyant}.
A refractive index matched experiment of particles in a turbulent duct flow was performed by \citet{zade2018experimental}.
Further experiments in various experimental facilities were reported, including MRI measurements in an annular Couette setup containing a dense granular suspension \cite{fall2010shear}, rheometer experiments of a colloidal suspension \cite{cwalina2014material}, refractive index matched experiments of a dense emulsion in pipe flow \cite{abbas2017pipe}, and MRI measurements in a particle-laden pipe flow \cite{leskovec2020pipe}. For the latter study a flattening of the velocity profile is reported in case of a bulk solid volume fraction of 0.2 and suspension Reynolds number ($Re_s = u_b D/\nu_s$ with $u_b$ the mixture bulk velocity and $\nu_s$ the suspension viscosity) of 700.

Recently, shear-induced migration in pipe flow has been reported by \citet{ardekani2018numerical}, who performed a numerical study of heat transfer in suspensions. In this study, the Reynolds numbers ($Re = u_b D/\nu$) were defined using the (uncorrected) continuous phase viscosity.
For the laminar case ($Re$ = 370) the velocity profiles are found to flatten for increasing volume fraction.
Moreover, in the turbulent region ($Re$ = 5300) a solid particle core is observed for higher volume fractions.
This is reported to be consistent with the findings of the study by \citet{lashgari2014laminar}, who reported an inertial shear-thickening regime for higher volume fractions. This regime is dominated by the particle induced stresses.
Similar behavior as reported by \citet{ardekani2018numerical} is also observed in a numerical study of a particle-laden channel flow by \citet{yousefi2021regimes}.


Experimental studies of suspension behavior in pipe flow are predominantly performed in the Stokes regime in order to avoid inertial effects, because these experiments are often devised in conjunction with theoretical analysis. Inertial effects significantly complicate this theoretical analysis.
However, in many natural and industrial processes inertial effects cannot be neglected. From the review above it is evident that the focus of the vast majority of studies was on the non-inertial regime. Open questions remain for higher Reynolds numbers, for instance the spatial distribution of particles and the effect thereof on the total pressure drop.
Reliable experimental data is of key importance for validation and development of theoretical models and numerical methods, in order to provide insight in the exact suspension dynamics and corresponding regimes.
Therefore, the main focus is on establishing a systematic, reliable data set. This is here achieved by a combination of experiments and numerical simulations.
A comparison is made using six cases, where each case has a unique combination of Reynolds number and bulk solid volume fraction. This allows to validate the experimental and numerical methods across different flow regimes, each exhibiting characteristic flow behavior. In contrast to the majority of previous studies, experiments are performed for higher Reynolds numbers.

\section{\label{Sect:experimentaldetails}Experimental details}
\subsection{\label{sect:experimentalsetup}Experimental facility}
\begin{figure}[b]
    \centering
    
    \subfloat[]{
    \includegraphics[width=0.66\textwidth]{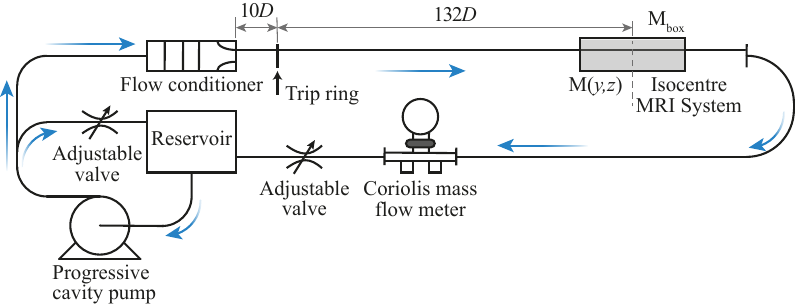}
    \label{fig:1a}}
    \hfill
    \subfloat[]{
    \includegraphics[width=0.31\textwidth]{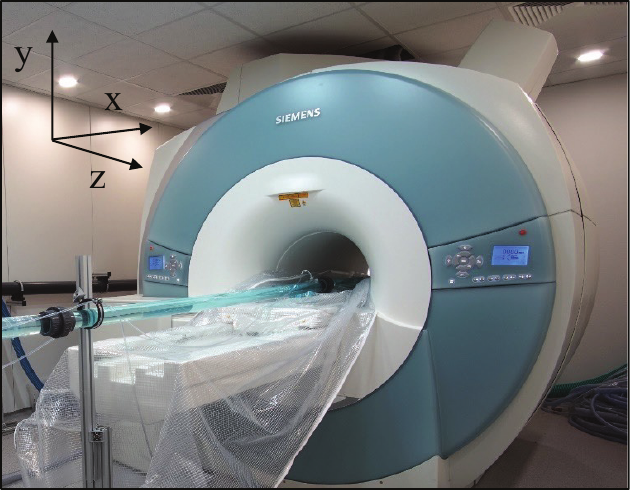}
    \label{fig:1b}}
    \caption{Schematic overview of the experimental setup \protect\subref{fig:1a} and a photograph of the MRI system and part of the experimental facility \protect\subref{fig:1b}. The Cartesian coordinate system used for data sampling is superimposed in the top left corner of the photograph, with the $z$-component in the streamwise direction.}
    \label{fig:1}
\end{figure}

A schematic of the experimental setup, including a photograph of the MRI system and part of the experimental setup is shown in Fig.~\ref{fig:1}. Experiments are performed in a 30.35 $\pm$ 0.12 mm inner diameter pipe. The suspension is transported using a progressive cavity pump (AxFlow B.V., Lelystad, the Netherlands). Using a settling chamber in combination with a smooth contraction, a laminar flow is maintained for $Re$ up to at least 3500. A concentric trip ring (inner diameter, $d_i$ = 25 mm), comparable with \citet{wygnanski1973transition}, is used to ensure a fixed transition for single-phase flows at $Re \approx$ 2000. MRI measurements are obtained at a distance of 132$D$ downstream of the orifice. A square box (inner dimensions: 0.1 $\times$ 0.1 $\times$ 0.4 m$^3$) containing a water-glycerol mixture is placed around the pipe at the isocentre of the scanner. The MRI signal of the fluid inside this box is used to account for the signal shift during experiments, this will be further discussed in Sect. \ref{MRI}. An inline Coriolis mass flow meter (KROHNE OPTIMASS 7050c) is used in the return loop to monitor the flow rate.
For average volume fractions exceeding 40 percent, the flowmeter caused jamming due to its narrower inner diameter.
Therefore, the bulk flow rate for all experiments is derived from the liquid velocity profile obtained using the MRI scanner, which proved to be in good agreement with the mass flow meter (average error less than 2\% for the single-phase and $\phi_b$ = 0.1 measurements; for the turbulent cases an error less than 1\%).
A temperature measurement is obtained in the reservoir using a PT100 temperature probe.
Furthermore, a cooling system (type: JULABO FT402) is used to minimize the temperature increase due to the heat addition from the pump.
The return loop is connected to a reservoir, closing the loop as the pump is fed from this reservoir.
In order to achieve low flow rates, a bypass is installed from the pump exit to the reservoir.
In combination with a valve, located in the main circuit (i.e., after the bypass), single-phase Reynolds numbers as low as 500 could be achieved.
For the particle-laden experiments, a mechanical stirrer is placed in the reservoir to keep the particles suspended. Small temperature variations lead to small rising or settling velocities of the particles, as the density of the fluid and particles are both sensitive to even the smallest temperature fluctuations \cite{dash2020particle}. A small residence time effect is observed on the solid volume fraction distribution. However, for the measurements presented in this study this effect was negligible (see also the axi-symmetry for all cases in Fig.~\ref{fig:4a}).\\

\noindent Unexpanded polystyrene particles (Synthos EPS; $d$ = 1.75 $\pm$ 0.12 mm; skewness $S$ = 0.698; $\rho$ = 1.032 $\pm$ 0.1\% kg/L) are used as dispersed phase. Note that there is approximately $\pm$ 7\% variation in the particle size, which has an important effect on the packing of particles in dense regions.
A glycerol-water mixture (mass ratio: 13.6/86.4\%) is used to obtain a density matched system.
The viscosity of the suspension, $\mu_s$, is determined a posteriori, based on the temperature of the suspension \cite{cheng2008formula}.
Furthermore, Eilers' viscosity correction \cite{eilers1941viskositat} is applied to correct for the increased viscosity due to the suspended particles. For the maximum packing fraction, $\phi_m$ = 0.64 is selected. Based on previous experimental findings this is found to be an appropriate choice for the used suspension \cite{hogendoorn2018particle, hogendoorn2021suspension}. Preparation of the suspension is based on the mass ratio of the particles and the glycerol-water mixture.
Starting with a single-phase system, particles are added in steps of 10\% up to a volume fraction of 50\%.
In addition, experiments are performed for a volume fraction of 45\%.
Moreover, copper sulfate (CuSO\textsubscript{4}; 1 g/L) is added to increase the T\textsubscript{1} decay, resulting in an enhanced signal-to-noise ratio of the MRI measurements. Note that the effect on density or viscosity of the original mixture will be negligibly small.\\

Practical limitations prevented the simultaneous measurement of the pressure drop along with the MRI measurements. Therefore, average pressure drop measurements were obtained in a separate series of experiments in the exact same flow loop for a range of $Re_s$ and $\phi_b$, using smaller increments spanning all cases. Subsequently, the pressure drops corresponding to each of the six studied MRI cases are determined using bi-linear interpolation in $Re_s$ and $\phi_b$. 

\subsection{\label{MRI}MRI system and settings}
The MRI system used is a MAGNETOM Trio 3T Whole-Body scanner (Siemens, Erlangen, Germany).
This scanner is located in the MRI laboratory at the Institute of Fluid Mechanics at the University of Rostock.
This laboratory is dedicated to study fluid mechanics applications, in contrast with the majority of other MRI facilities.
The measurement parameters used for the experiments are shown in Table \ref{table:1}.
The MRI system has a maximum gradient amplitude of 40 mT/m and a maximum gradient slew rate of 200 T/m/s. Two standard receive-only body coils were used to receive the signal.

\begin{table}[htb]
	\centering
	\caption{MRI parameters used}
	\begin{tabular}[t]{ll}
	\toprule
	\textbf{Parameter}              &\textbf{Value}\\
	\midrule
	Matrix size                     & 1 $\times$ 640 $\times$ 640         \\
	Non-isotropic resolution        & 50 $\times$ 0.3 $\times$ 0.3 mm$^3$         \\
	                                & 28.6 $d~\times$ 0.17$d~\times$ 0.17$d$ \\
	Repetition time (TR)            & 22 ms         \\
	Echo time (TE)                  & 9 ms         \\
	RF flip angle                   & 5$^{\circ}$         \\
	Receiver bandwidth              & 280 Hz/pixel         \\
	Velocity Encoding               & 0.1 \textendash 1.7 m/s         \\
	Number of samples               & 32         \\
	Total acquisition time          & 30 min for each combination of $Re_s$ and $\phi_b$        \\
	\bottomrule
	\end{tabular}
	\label{table:1}
\end{table}

For the MRI measurements a non-isotropic spatial resolution is used: 50 $\times$ 0.3 $\times$ 0.3 mm$^3$ in the $x$, $y$, and $z$ direction, respectively.
This non-isotropic resolution behaves similar to a spatial average along the $x$, or streamwise direction.
The $y$-$z$ plane is perpendicular to this streamwise direction, with $y$ the vertical and $z$ the horizontal component (see also the coordinate system in Fig.~\ref{fig:1b}).

In the current suspension consisting of glycerol-water mixture and polystyrene particles, only the liquid phase creates a signal. The presence of particles reduces the amount of liquid within a voxel, resulting in a lower MRI signal magnitude $M(y,z)$. This magnitude can therefore be used to quantify the local volume fraction. The time-averaged particle volume fraction $\phi(y,z)$ is reconstructed from the signal magnitude of a particle-laden flow measurement and a reference measurement, $M_{ref}(y,z)$, without particles in the flow:
\begin{equation}
    \phi(y,z) = 1 - \frac{M(y,z)/\overline{M_{box}}}{M_{ref}(y,z)/\overline{M_{box,ref}}},
    \label{eq:1}
\end{equation}
where $\overline{M_{box}}$ is the average signal magnitude in the glycerine-filled box around the pipe (see Fig. \ref{fig:1}). This correction improves the measurement accuracy since the two measurements can be taken hours apart and the signal level may have changed slightly during that time. This magnetic drift (or $B_0$ fluctuation) is a common, mostly unavoidable issue in MRI recordings.

Note that the actual and reference measurements are performed with identical settings to avoid other effects that may influence the signal magnitude. Except for the particle volume fraction, the flow conditions are the same for the two cases, i.e., same experimental setup with the same total volumetric flow rate.
For particle concentration and velocity measurements, two different MRI protocols are used: the particle \textit{concentration} measurements are performed with a velocity-compensated MRI acquisition sequence to decrease the sensitivity of the signal magnitude to changes in the velocities between the particle-laden case and the reference case. \textit{Velocity} measurements are conducted using one-directional Phase-Contrast MRI \cite{pelc1991encoding}. Except for the encoding technique, all measurement parameters are identical for the two MRI protocols. Note that the velocities shown in this study are the intrinsic fluid velocities, as only the fluid phase provides a measurable signal.

The velocity and solid volume fraction results obtained on the Cartesian grid are interpolated on a non-equidistant cylindrical grid using bi-linear interpolation in ($y,z$) for ($r,\theta$). A non-equidistant grid is used, as the grid density increases for increasing $r$ in order to correct for the higher information density in the direction of the wall. Subsequently, the data are averaged over the azimuthal direction, $\theta$, in which the flow is statistically homogeneous, to obtain the mean solid fraction as function of the radial coordinate, $r$. Comparison of these integrated radial velocity profiles with the original Cartesian data results in an error less than 1\% for the bulk velocity (across all cases). This value comprises all errors of this interpolation, including the accuracy of the determination of the pipe centre etc. In order to keep the results consistent, the bulk solid volume fraction was also obtained in the cylindrical coordinate system using:
\begin{equation}
    \phi_b = \frac{2}{R^2}\int_{r=0}^{r=R} \phi(r)rdr.
    \label{eq:2}
\end{equation}

\begin{figure}[ht]
    \centering
    \includegraphics[width=0.5\textwidth]{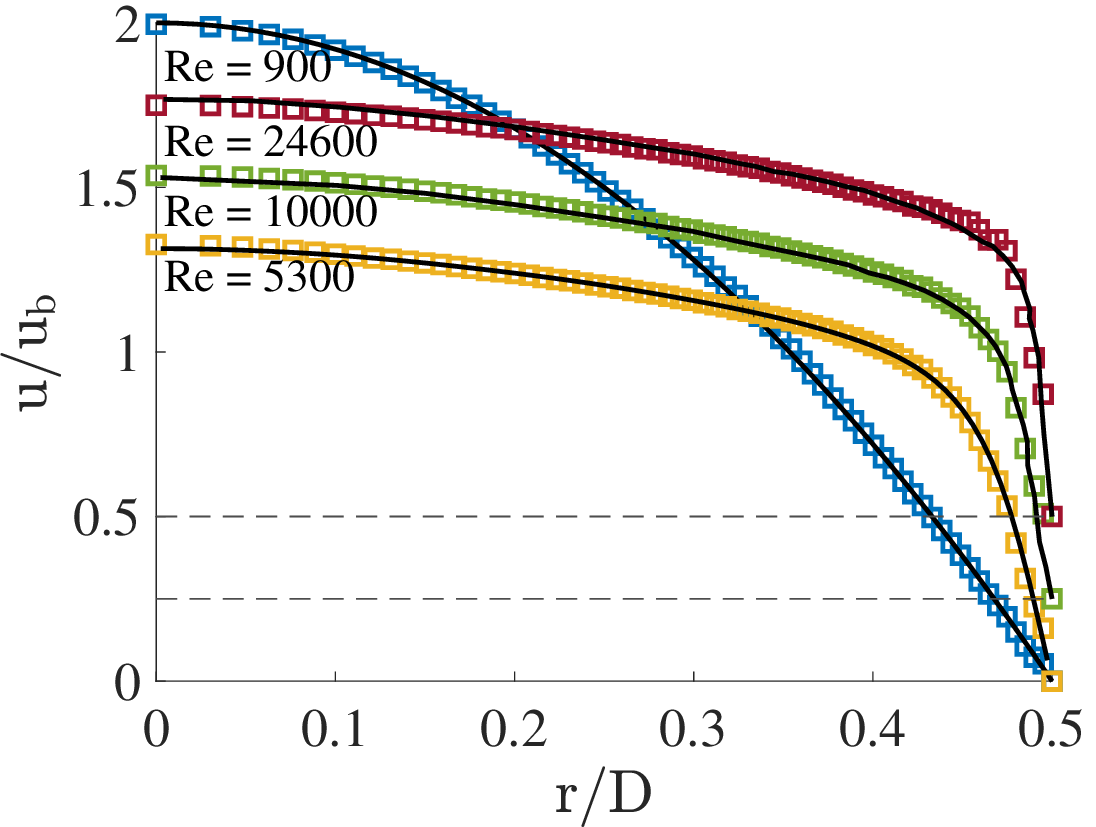}
    \caption{Single-phase velocity profiles (markers) compared to reference data (solid lines) for $Re \approx$ 900, 5300, 10~000, and 24~600. For clarity, the cases corresponding to $Re$ = 10~000 and 24~600 are shown with a vertical offset of 0.25 and 0.5, respectively.}
    \label{fig:2}
\end{figure}

\noindent First, the MRI results for single-phase flow at $Re$ = 900, 5~300, 10~000, and 25~000 are validated with reference data from literature \cite{eggels1994fully, den1997reynolds}. The results of this single-phase validation are shown in Figure \ref{fig:2}; note that an offset is used for the two highest $Re$. The comparison is restricted to the mean streamwise velocity profile as function of radius, as this is the only quantity obtained from our MRI measurements.
For all cases an average error between the profiles of less than 1\% of the bulk velocity is found. This confirms that the current MRI protocol can measure both laminar and turbulent profiles accurately.

Also, the error in the concentration measurement is determined using error propagation \cite{bruschewski2016estimation}.
This error decreases for increasing number of repetitions; 32 repetitions per measurement are found to result in a maximum error (STD/mean) of 2.3\% across all measurements.

For the highest volume fraction experiment ($Re_s$ = 1732, $\phi_b$ = 0.466) the absolute discrepancy between the target volume fraction, $\phi$ = 0.45, and measured volume fraction, $\phi_b$ = 0.466, by the MRI scanner is found to be 0.016. Note that this value comprises all kinds of errors. These errors can occur due to dead spots in the experimental facility, reconstruction errors, slight difference in magnetic susceptibility of particles and fluid, higher orders of motion (see, e.g., \citet{schmidt2021reynolds}), etc. Moreover, due to particle aggregation in the pipe centre, the effective volume fraction in the pipe deviates from the solid volume fraction in the feeding reservoir. As the velocity in the centre is highest, the particles are convected faster than the bulk flow. This is similar to the Fåhræus effect, where red blood cells aggregate in the vessel centre in the microcirculation \cite{pries2008blood}.
Therefore, the analysis in this study is based on the bulk solid volume fraction resulting from circumferential integration of the measured particle volume fraction profile (Eq.~\ref{eq:2}). This will result in a better estimate of the actual average solid volume fraction in the measurement section, and thereby also in a more accurate determination of the suspension Reynolds number.

\subsection{\label{cases}Selected cases to compare to DNS}
Six cases are selected from the experiments, which are compared with direct numerical simulations.
These cases are selected for various combinations of $Re_s$ and $\phi_b$, in order to shed light on suspension flow dynamics under different conditions.
Simultaneously, this allows to validate the obtained experimental and numerical data set for a range of flow conditions.
The six cases with the corresponding experimental conditions are listed in Table~\ref{table:2}.
The variation in fluid viscosity, $\nu$, is due to temperature variations of the suspension. In particular for the higher velocities and solid volume fractions, more heat from the pump is added to the suspension.

From the experiments the time-averaged intrinsic liquid velocity, $u_l(r)$, and particle volume fraction profiles (or void fraction profiles, defined as one minus the bulk solid volume fraction) are available.
From this information, the intrinsic liquid bulk velocity, $u_{l,b}$ is determined, being defined as:
\begin{equation}
    u_{l,b} = \frac{\int_{r=0}^{r=R} u_l(r) \, (1-\phi(r)) \, 2 \pi r \, dr}{\int_{r=0}^{r=R} (1-\phi(r)) \, 2 \pi r \, dr}.
    \label{eq:3}
\end{equation}
This represents a weighted average using the local void fraction. 
The corresponding liquid bulk Reynolds number is defined as $Re_l = u_{l,b} D/\nu$. We also determined the ratio $u_b/u_{l,b}$ with $u_b$ the bulk mixture velocity. The latter has not been measured directly, but was estimated from Eq.~\ref{eq:3} with $1-\phi$ put equal to 1. This is a valid approach provided that the macroscopic slip velocity between the liquid and the solid phase is zero or $\phi(r) \, u_s(r) + (1-\phi(r)) \, u_l(r) \approx u_l(r)$; a small bias may be expected from wall slip of the solid phase at the pipe wall. This is also confirmed by the DNS data. The bulk mixture/liquid velocity ratio can be used as a metric which reveals information about the solid volume fraction distribution and the corresponding flow regime. See for instance Case 3, where an almost uniform volume fraction distribution is observed (Fig.~\ref{fig:6c}) for $u_b/u_{l,b} = 1.01$ and thus close to unity. In Table~\ref{table:2} the suspension Reynolds number is based on the bulk mixture velocity and the suspension viscosity, thus $Re_s = Re_l (u_b/u_{l,b})/(\nu_s/\nu)$. Finally, $f$ represents the Darcy-Weisbach friction factor defined by $f \equiv \Delta p/(\rho u_{b}^2) \cdot 2D/L$.

\begin{table}[htb]
	\centering
	\caption{Selected cases and corresponding flow conditions and parameters for the experiments. Explanation of the different columns is provided in the text.}
	\begin{tabular}[t]{p{1.8cm} p{1.8cm} p{1.8cm} p{1.8cm} p{1.8cm} p{1.8cm} p{1.8cm} p{1.8cm} p{1.8cm}}
	\toprule
	\textbf{Case}   &\textbf{$Re_l$}  &\textbf{$\phi_b$}  &\textbf{$\nu$ [$m^2/s$]} &\textbf{$u_{l,b} [m/s]$}  &\textbf{$\nu_s/\nu$} &\textbf{$u_{b}/u_{l,b}$}    &\textbf{$Re_s$} & \textbf{$f$}\\
	\midrule
	1               & 2339       & 0.252             & 1.440 E--6      & 0.111      & 2.31      & 1.070    & 1083  & 6.50 E--2\\
	2               & 6012       & 0.272             & 1.443 E--6      & 0.286      & 2.53      & 1.067    & 2532  & 4.40 E--2\\
	3               & 7253       & 0.089             & 1.436 E--6      & 0.343      & 1.28      & 1.010    & 5740  & 3.99 E--2\\  
    4               & 9977       & 0.195             & 1.436 E--6      & 0.472      & 1.82      & 1.025    & 5613   & 3.59 E--2\\ 
	5               & 15270      & 0.315             & 1.420 E--6      & 0.715      & 3.15      & 1.050    & 5096   & 3.05 E--2\\  
    6               & 16126      & 0.466             & 1.287 E--6      & 0.684      & 9.89      & 1.062    & 1732   & 4.32 E--2\\ 
	\bottomrule
    \end{tabular}
	\label{table:2}
\end{table}

\section{\label{Sect:DNS}Simulation details}

\subsection{\label{Goveq}Governing equations}
In the particle-resolved direct numerical simulations the full Navier-Stokes equations are solved for the incompressible carrier flow and the Newton-Euler equations for the motion of the particles. Using the particle diameter ($d$) and intrinsic liquid bulk velocity ($u_{l,b}$) as characteristic scales for the normalisation, the non-dimensional Navier-Stokes equations read:
\begin{subequations}
\begin{eqnarray}
\nabla \cdot \bf{u} & = & 0 \, , \label{vgl1a} \\
\frac{\partial \bf{u}}{\partial t} + \nabla \cdot \bf{uu} & = & -\nabla p_e -\nabla p + \frac{D}{d} \, \frac{1}{Re_l} \, \nabla^2 \bf{u} \, ,  \label{vgl1b}
\end{eqnarray}
\end{subequations}
where $dp_e/dx$ is the externally imposed pressure gradient by which the flow is driven, and $p = p_t - p_e(x) - \rho {\bf g} \cdot {\bf x}$ is a modified pressure with $p_t$ the total pressure and $\bf g$ the gravitational acceleration. The non-dimensional Newton-Euler equations for the particle linear (${\bf u}_c$) and angular ($\text{\boldmath$\omega$}_c$) velocity are given by:
\begin{subequations}
\begin{eqnarray}
\frac{\pi}{6} \frac{d {\bf u}_c}{dt} & = & \oint_{A_p} \left( \text{\boldmath$\tau$} \cdot {\bf n} \right) \, dA - \frac{\pi}{6} \, \nabla p_e + {\bf F}_c \, , \label{vgl2a} \\
\frac{\pi}{60} \frac{d \text{\boldmath$\omega$}_c}{dt} & = & \oint_{A_p} {\bf r} \times ({\bf \text{\boldmath$\tau$}} \cdot {\bf n} ) \, dA + {\bf T}_c \, , \label{vgl2b} 
\end{eqnarray}
\end{subequations}
where it is used that the particles are neutrally buoyant and spherical, $A_p$ denotes the particle surface, $\text{\boldmath$\tau$} = -p {\bf I} + \frac{D}{d} \, \frac{1}{Re_l} \, \left( \nabla {\bf u} + \nabla {\bf u}^T \right)$ is the fluid stress tensor with ${\bf I}$ the unit tensor, ${\bf n}$ is the outward unit normal on the particle surface pointing into the fluid phase, and ${\bf F}_c$ and ${\bf T}_c$ are the collision force and torque, respectively. The Navier-Stokes and Newton-Euler equations are coupled with each other through the no-slip/no-penetration condition at the surface of the particles: 
\begin{eqnarray}
{\bf u} & = & {\bf U}_p \;\;\; \textrm{at } A_p \, , \label{vgl3}
\end{eqnarray}
where ${\bf U}_p = {\bf u}_c + \text{\boldmath$\omega$}_c \times {\bf r}$ is the local particle velocity on the particle surface.

\subsection{\label{Nummeth}Numerical approach}
The present DNS makes use of the computationally efficient Immersed Boundary Method (IBM) of \citet{breugem2012} for the particle-fluid coupling, which is a modified version of the original IBM proposed by Uhlmann \cite{Uhlmann2005}. In this approach, a Eulerian (fixed) and uniform 3D grid is used in which the particles are `immersed'. The interface condition (Eq.~\ref{vgl3}) is enforced by good approximation by means of locally adding forces, ${\bf f}_{IB}$, to the right-hand side of Eq.~\ref{vgl1b} in a spherical shell around the particle/fluid interface. The computation of the IBM force is embedded as an additional step in the fractional step (predictor--corrector) scheme used to integrate the Navier-Stokes equations. To this purpose, on every particle a uniform Lagrangian grid on the particle surface is employed that moves along with the particle. Using a regularized delta function \cite{roma1999}, the provisional fluid velocity obtained from solving Eq.~\ref{vgl1b} is interpolated from the Eulerian to the Lagrangian grid (${\bf u}^{*} \rightarrow {\bf U}^*$), then the IBM force on the Lagrangian grid is computed from the apparent particle-fluid slip velocity (${\bf F}_{IB} = ({\bf U}_p - {\bf U}^*)/\Delta t$, with $\Delta t$ the computational time step), following which the computed force is interpolated back to the Eulerian grid (${\bf F}_{IB} \rightarrow {\bf f}_{IB}$). Finally, the provisional velocity is corrected for the presence of the particles (${\bf u}^* \rightarrow {\bf u}^* + \Delta t \, {\bf f}_{IB}$). To overcome the problem of overlapping interpolation kernels of neighbouring Lagrangian grid points, a multidirect forcing scheme \cite{luo2007} is implemented in which the computation of the IBM force is improved by a few iterations (six in the present DNS). The accuracy of the IBM is further enhanced by slight inward retraction of the Lagrangian surface grid by $0.3 \Delta x$. Main advantages of the IBM are that no regridding is required when particles move in space and that the uniform grid allows for the use of fast solvers for, e.g., the Poisson equation for the correction pressure. \\
In the DNS a rectangular computational domain and a Cartesian fluid grid are used, which do not conform to the cylindrical flow geometry. Hence, another efficient IBM is used to enforce the no-slip/no-penetration condition on the pipe wall. This IBM is similar to the volume-penalization method described in \citet{breugem2014} and is incorporated as an additional step in the fractional step scheme. Directly after the first provisional velocity is obtained from Eq.~\ref{vgl1b}, the provisional velocity is multiplied with the cell pipe volume fraction, ${\bf u^*} \rightarrow \beta_{P} \, {\bf u}^*$ for $r \leq R + 2\Delta x$. The pipe volume fraction within a grid cell, $\beta_P$, is computed once at the start of a simulation from a level-set approach based on the signed distance of the eight cell corners to the pipe wall \cite{kempe2012,breugem2012}. Note that the volume penalization method results in a smooth pipe boundary with a radial thickness of $O(\Delta x)$, but as long as $\Delta x/R \ll 1$ the effect on the overall flow behavior can be kept sufficiently small. \\
In the fractional step scheme, first a provisional velocity is computed from Eq.~\ref{vgl1b}, which is then corrected for the presence of the pipe wall, subsequently corrected for the presence of the particles, and finally followed by a pressure-correction step to enforce Eq.~\ref{vgl1a}. An efficient direct (FFT-based) solver is used for solving the Poisson equation for the correction pressure. The coupling of the Navier-Stokes with the Newton-Euler equations is explicit/weak: given the positions and velocities of the particles from the previous time step, first the Navier-Stokes equations are integrated in time and then the Newton-Euler equations from the computed IBM force distribution on the Lagrangian particle grid. \\
The Navier-Stokes and Newton-Euler equations are both integrated in time with a 3-step Runge-Kutta (RK) method. The time step, $\Delta t$, is chosen sufficiently small to ensure numerical stability. The Navier-Stokes equations are discretized in space on a staggered uniform Cartesian grid with the finite-volume method. Spatial derivates are computed from the central-differencing scheme. The computational domain is a rectangular box with walls at the bottom and top. Periodic boundary conditions are imposed in the horizontal (streamwise and spanwise) directions, while a no-slip/no-penetration condition is imposed at the walls in the vertical direction. The dimensions of the box are chosen such that pipe fits within the box with a distance of at least 2 grid cells between the pipe wall and the outer domain boundaries as required by the volume penalization IBM. The flow is forced by adjusting the streamwise pressure gradient, $dp_e/dx$, such that the intrinsic liquid bulk velocity ($u_{l,b}$) is maintained at a constant value. The pressure gradient is iteratively adjusted in the first RK step in the multidirect forcing scheme mentioned before, and the total pressure gradient is subsequently fixed in the second and third RK step. Test simulations of single-phase pipe flow for $Re=1000$ and $5300$ using $\Delta x/R = 7.25 \cdot 10^{-3}$ (same resolution as used in particle-laden flows) showed excellent agreement of the obtained Darcy-Weisbach friction factor with $64/Re$ for laminar flow and the Blasius' correlation for turbulent flow in a hydraulically smooth pipe, respectively. For both Reynolds numbers, the error amounts less than 0.4\%.

\subsection{\label{collision}Collision model}
Particle collisions are modelled with the frictional soft-sphere collision model described in \citet{costa2015}. This is a linear spring-dashpot model in which the rigid particles are allowed numerically to slightly overlap each other. The normal force acting on particle $i$ from a collision with particle $j$, is computed from the normal overlap along the line of centers ($\text{\boldmath$\delta$}_{ij,n}$) and the relative normal particle velocity (${\bf u}_{ij,n}$) as:
\begin{subequations}
\begin{eqnarray}
{\bf F}^c_{ij,n} & = & -k_n \text{\boldmath$\delta$}_{ij,n} - \eta_n {\bf u}_{ij,n} \, , \label{vgl5a}
\end{eqnarray}
where $k_n$ and $\eta_n$ are the normal spring and dashpot coefficients, respectively. \\
The tangential component of the collision force is modelled in a similar way, but undergoes a stick-slip transition when the tangential force exceeds a threshold value dependent on the normal force component:
\begin{eqnarray}
{\bf F}^c_{ij,t} & = & \textrm{min} \left( \, \lVert -k_t \text{\boldmath$\delta$}_{ij,t} - \eta_t {\bf u}_{ij,t} \rVert \, , \, \lVert - \mu_c {\bf F}^c_{ij,n} \rVert) \, \right) \, {\bf t}_{ij} \, , \label{vgl5b}
\end{eqnarray}
where $k_t$ and $\eta_t$ are the tangential spring and dashpot coefficients, respectively, $\mu_c$ is the Coulomb coefficient of sliding friction, and ${\bf t}_{ij}$ is a unit vector pointing in the direction of the test force for the stick regime:
\begin{eqnarray}
{\bf t}_{ij} & = & \frac{-k_t \text{\boldmath$\delta$}_{ij,t} - \eta_t {\bf u}_{ij,t}}{\lVert -k_t \text{\boldmath$\delta$}_{ij,t} - \eta_t {\bf u}_{ij,t} \rVert} \, . \label{vgl5c}
\end{eqnarray}
The collision torque is computed from the collision force as:
\begin{eqnarray}
{\bf T}^c_{ij} & = & a \, {\bf n}_{ij} \times {\bf F}^c_{ij,t} \, , \label{vgl5d}
\end{eqnarray}
\end{subequations}
where ${\bf n}_{ij}$ is the unit vector along the line of centers pointing from particle $i$ to particle $j$. \\
The normal and tangential spring and dashpot coefficients can be expressed in terms of the dry normal ($e_{n,d}$) and tangential ($e_{t,d}$) coefficients of restitution and the collision time duration \cite{costa2015}. The latter is typically set equal to a few computational time steps ($N_c \Delta t$) as to sufficiently resolve the collisions in time ($N_c = 8$ in the present DNS). \\
Particle-wall collisions are modelled by treating the pipe wall as a spherical particle with infinite radius. The total collision force and collision torque acting on a particle are computed from the sum over all contributions from the particles/wall with which it is in contact. Numerically, the collision force and torque are integrated in time with the Crank-Nicolson scheme and two iterations are used to determine the force and the torque at the new time level as function of the new particle velocities and positions. In addition, to further improve the temporal accuracy, subintegrations are performed for the collisions with the number of substeps equal to 40 in the present DNS and equidistantly divided over the total time step comprising of three RK steps. \\
Finally, the model also includes corrections on the lubrication force that a particle experiences when it is in close proximity with another particle at a distance less than a grid cell between their interfaces. Because a fixed Cartesian grid is used, the DNS then lacks sufficient grid resolution in the intervening gap to resolve the local flow between the nearby particles. The corrections are applied only for the normal lubrication force between nearby particles and for particles close to the pipe wall. They are based on asymptotic expansions of exact analytical solutions \cite{dance2003}. The diverging behavior of the expansions is capped when the gap width drops below a threshold value associated with particle/wall roughness \cite{costa2015}. The threshold distance is set equal to $2 \cdot 10^{-3} \, d$ in the present simulations. In the DNS code, the lubrication force corrections are added to the collision force and are hence integrated in the same manner.

\subsection{\label{compparameters}Flow parameters and computational settings}
The DNS code is written in modern Fortran with the MPI extension for parallel computing on multi-core systems with distributed memory. The parallelization consists of a 2D pencil decomposition of the computational domain and a controller/worker technique for handling the particles. The present DNS was conducted on the Dutch National Supercomputer Snellius on 128 CPU cores for about 2-3 weeks per case.

The main input parameters for the DNS are the bulk liquid Reynolds number ($Re_l$), the bulk solid volume concentration ($\phi_b$) and the particle-to-pipe diameter ratio ($d/D$). Their values were chosen based on the experiments. Table~\ref{table:2b} contains the values of $Re_l$ and $\phi_b$ for all six cases investigated, while $d/D$ is fixed at 0.058. Note that slight differences are present between the experiments (Table~\ref{table:2}) and the DNS (Table~\ref{table:2b}); the input for the DNS was based on a preliminary analysis of the experimental data, which in some cases differed a bit from the final analysis of the data. Perhaps most notable is the difference in concentration in Case 5: $\phi_b = 0.315$ in the experiments versus $0.310$ in the DNS. Overall, the differences are minor and, also considering the experimental uncertainty, deemed negligible. \\
The parameters of the collision model are fixed at $e_n = 0.97$, $e_t = 0.1$ and $\mu_c = 0.39$, except for case 6$^*$ where $\mu_c = 0$ (i.e., frictionless particles and pipe wall) was set to investigate the effect of interparticle friction on the flow dynamics in this case. The value of $\mu_c=0.39$ was taken from recent tilted-flume experiments on immersed polystyrene beads with a similar diameter but from a different supplier \cite{shajahan2023}. The value of $e_n$ has been reported frequently in literature for different materials, while the current value of $e_t$ was based on an educated guess as to promote early transition from stick to slip behavior.

\begin{table}[htb]
	\centering
	\caption{Flow conditions and parameters in DNS for all six cases investigated. Explanation of the different columns is provided in the text. Case 6$^*$ is the same as case 6 but with the Coulomb coefficient of sliding friction set equal to 0.}
	\begin{tabular}[t]{p{1.cm} p{1.4cm} p{1.2cm} p{1.3cm} p{1.3cm} p{1.cm} p{1.3cm} p{1.5cm} p{1.1cm} p{1.1cm} p{0.8cm} p{1.4cm}}
	\toprule
	\textbf{Case} &\textbf{$Re_l$} &\textbf{$\phi_b$} &\textbf{$u_{b}/u_{l,b}$} &\textbf{$u_{s,b}/u_{l,b}$} &\textbf{$\nu_s/\nu$}  &\textbf{$Re_s$} &\textbf{$f$} 
	&\textbf{$\Delta x_s^+$} &\textbf{$\Delta t_s^+$} &\textbf{$N_s$} &\textbf{$T_s \, u_{b}/R$}\\
	\midrule
        1   & 2346.3    & 0.2520    & 1.065  & 1.261  & 2.31  & 1080.6    & 8.90 E--2 & 0.41 & 0.015 & 260   & 161.6\\
        2   & 6038.2    & 0.2721    & 1.068  & 1.252  & 2.53  & 2541.0    & 7.31 E--2 & 0.88 & 0.027 & 217   & 126.9\\
        3   & 7323.6    & 0.0900    & 1.008  & 1.094  & 1.28  & 5774.4    & 4.86 E--2 & 1.63 & 0.037 & 291   & 302.8\\  
        4   & 10045.6   & 0.1948    & 1.034  & 1.174  & 1.82  & 5690.5    & 5.62 E--2 & 1.73 & 0.036 & 397   & 177.9\\ 
        5   & 14778.6   & 0.3100    & 1.069  & 1.224  & 3.07  & 5138.6    & 7.74 E--2 & 1.83 & 0.046 & 220   & 203.2\\  
        6   & 16269.8   & 0.4661    & 1.062  & 1.133  & 9.88  & 1739.4    & 26.25 E--2 & 1.14 & 0.033 & 267   & 76.4\\ 
        6$^*$ & 16269.8 & 0.4661    & 1.051  & 1.109  & 9.88  & 1721.1    & 17.59 E--2 & 0.93 & 0.022 & 259   & 73.3\\
	\bottomrule
    \end{tabular}
	\label{table:2b}
\end{table}

The computational domain has dimensions $L_{x}\times L_{y}\times L_{z}$ of $87\,d \times 18\,d \times 18\,d$ in the streamwise, spanwise and vertical direction, respectively. A uniform Cartesian grid is used of 1392$\, \times \,$288$\, \times \,$288 grid cells. The flow in the immediate vicinity of the particles is thus resolved at a resolution of $d/\Delta x = 16$. This corresponds to 746 Lagrangian grid cells uniformly distributed over the surface of the spheres. For the chosen ratio $d / D$, we have 275.9 grid cells over the pipe diameter and a pipe length of $L/D$ = 5.05.

The flow in the DNS was initialized by imposing a laminar Poiseuille velocity profile in the pipe for both fluid and particles. The angular velocity of the particles was set to half the local vorticity in order to start from a smooth initial condition. Except for Case 6, the particles were initially randomly placed inside the pipe with the requirement of no overlap between the particles or with the pipe wall. This turned out not to be possible for Case 6: because the particles are injected one after the other and cannot move anymore once they have been injected, the average spacing between the injected particles is too high to reach concentrations well beyond 30\%. Hence for this case first a separate simulation was run in which the particles were forced to settle under gravity in a sufficiently long, closed, vertical pipe, and where only the Newton-Euler equations were solved with a simple model for the hydrodynamic drag. The packed bed thus obtained, was then placed in the actual pipe and the axial location of the particles multiplied with a constant factor such that the stretched packing filled the entire pipe. Case 6$^*$ was initiated by continuing the simulation of Case 6 with the Coulomb coefficient of sliding friction set to 0. The DNS was run for sufficiently long time in order to obtain a statistically fully developed flow (no trends in the streamwise pressure gradient, average streamwise particle velocity, etc.). It was then further run for a time $T_s$ to sample the flow for the statistical analysis detailed in section \ref{postproc}.

Table~\ref{table:2b} contains various output parameters obtained from the DNS. The velocity ratio $u_b/u_{l,b}$ is generally in good agreement with the estimates obtained from the MRI measurements with the biggest deviation seen for Case 5. $u_{s,b}/u_{l,b}$ is the solid/liquid bulk velocity ratio and a measure for the degree of core peaking of the solid volume fraction profile; it is highest for Cases 1 and 2. The viscosity ratio $\nu_s/\nu$ was obtained from Eilers' correlation with the same $\phi_m = 0.64$ as used for the analysis of the experiments. Note that the suspension viscosity depends on the local concentration and thus actually varies across the pipe. Furthermore, it is well known that at high concentration the suspension viscosity is sensitive to the degree of interparticle friction, being higher for higher values of $\mu_c$ \cite{guazzelli2018rheology}. However, a detailed assessment of the suspension rheology is out of scope of the present study and left for future research. The difference in the Darcy-Weisbach friction factor $f$ for Cases 6 and 6* clearly shows a strong effect of interparticle friction on the overall flow dynamics. The friction factor $f$ obtained from the DNS is in all cases larger than the experimental values, and for the most dense cases (5, 6 and 6*) even much larger. We will discuss this later (see the final paragraph in Sect.~\ref{Sect:comparison}). The non-dimensional grid spacing, $\Delta x_s^+ = \Delta x \, u_{\tau}/\nu_s$, is below 2 in all cases and deemed sufficiently small in case the flow would be a classical single-phase turbulent pipe flow. Also the time step normalised with the viscous time unit, $\Delta t_s^+ = \Delta t \, u_{\tau}^2/\nu_s$, is sufficiently small and the flow is thus considered well resolved in time in all cases. Finally, the number of statistical samples used, $N_s$, and the non-dimensional time interval, $T_s u_{b}/R$, over which the samples were taken, are listed in the last two columns of Table~\ref{table:2b}.

\subsection{\label{postproc}Postprocessing of DNS data}
The average intrinsic velocity profile of the liquid was obtained as follows. First, the {\it{superficial}} instantaneous volume average was computed from:
\begin{subequations}
\begin{eqnarray}
\langle u_l \rangle (y,z,t) & = & \frac{1}{L_x} \int_{0}^{L_x} \gamma_l(x,y,z,t) \, u_l (x,y,z,t) \, dx \, , \label{vgl7a}
\end{eqnarray}
where the brackets $\langle .. \rangle$ denote the volume average and $\gamma_l = 1 - \gamma_s$ with $\gamma_s$ the local solid volume fraction in a computational grid cell. The latter was computed from the positions of the particles in the same manner as the local pipe volume fraction was computed in the simulations. The instantaneous solid volume concentration was obtained from:
\begin{eqnarray}
\phi (y,z,t) & = & \frac{1}{L_x} \int_{0}^{L_x} \gamma_s(x,y,z,t) \, dx \, . \label{vgl7b}
\end{eqnarray}
The volume-averaged velocity and concentration distributions were then averaged over the number of samples taken:
\begin{eqnarray}
& & \overline{\langle u_l \rangle} (y,z) = \frac{1}{N_s} \sum_{q} \langle u_l \rangle ({y,z,t_q}) \;\;\; \textrm{and} \;\;\; \overline{ \phi } (y,z) = \frac{1}{N_s} \sum_q \phi ({y,z,t_q}) \; . \label{vgl7c} 
\end{eqnarray}
The averages are now still defined on the Cartesian grid ($y$,$z$). Similar to the processing of the MRI data, bilinear interpolation was used to interpolate them to a cylindrical grid ($r$,$\theta$). The resolution of this grid was chosen such that $\Delta r = \Delta x$ and $R \, \Delta \theta = \Delta x$. Next, the distributions were averaged over the $\theta$-direction:
\begin{eqnarray}
& & \overline{\langle u_l \rangle} (r) = \frac{1}{2 \pi/\Delta \theta} \sum_q \overline{\langle u_l \rangle} ({r,\theta_q}) \;\;\; \textrm{and} \;\;\; \overline{\phi} (r) = \frac{1}{2 \pi/\Delta \theta} \sum_q \overline{\phi } ({r,\theta_q}) \, . \label{vgl7d}
\end{eqnarray}
Finally, the {\it{intrinsic}} (phase-averaged) time and volume-averaged liquid velocity was obtained from:
\begin{eqnarray}
\overline{\langle u_l \rangle^l} & = & \frac{\overline{\langle u_l \rangle}}{1 - \overline{\phi}} \, . \label{vgl7g} 
\end{eqnarray}
\end{subequations}
To simplify the notation in the discussion of the results below, we will represent $\overline{\langle u_l \rangle^l}$ and $\overline{\phi}$ by $u_l$ and $\phi$, respectively. 

\section{\label{Sect:results}Results and discussion}
\subsection{\label{Sect:3DFields}Instantaneous 3D flow fields}
\begin{figure}[ht]
    \includegraphics[width=\textwidth]{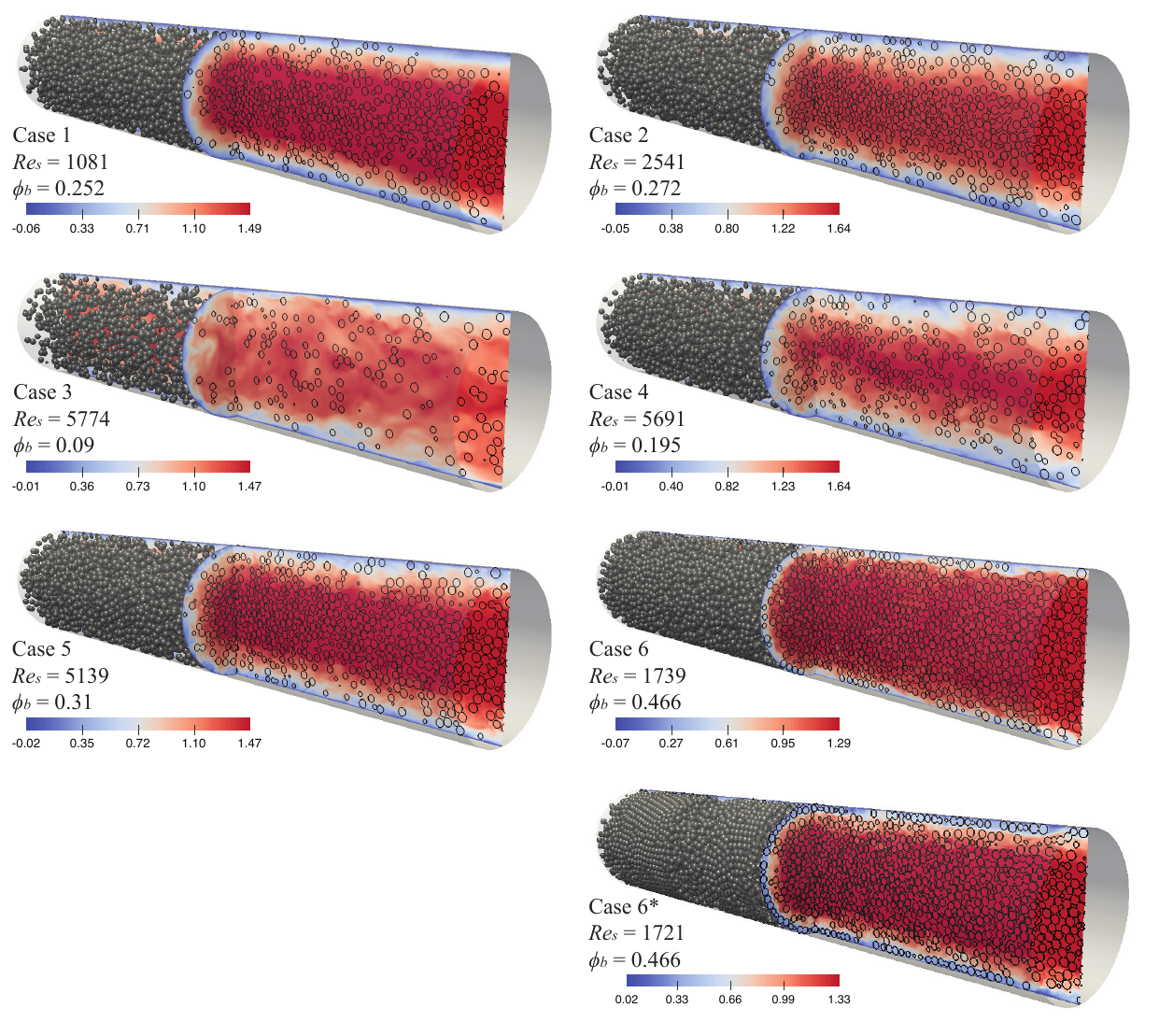} 
    \caption{Instantaneous flow snapshots obtained from DNS for all cases studied. The color denotes the local streamwise velocity in the fluid and solid phase, normalised with the intrinsic liquid bulk velocity. Contours indicate the particle positions.}
    \label{fig:3}
\end{figure}

Fig.~\ref{fig:3} shows instantaneous snapshots of the flow obtained from the DNS for all investigated cases. The color denotes the local streamwise velocity normalised with the intrinsic liquid bulk velocity ($u/u_{l,b}$ with $u$ the local fluid or solid phase velocity). The spheres and black contours indicate the particle positions. Cases 1, 2 and 5 appear very similar to each other, all displaying strong aggregation of the particles in the pipe core. This is also reflected in the similar high values for $u_{s,b}/u_{l,b}$ in these cases, see Table~\ref{table:2b}. Cases 3, 4, and 5 show a change from a homogeneous to a core-peaking distribution of the solid volume fraction for increasing $\phi_b$ at approximately constant $Re_s$. In Case 3, with the lowest bulk solid volume fraction of 0.09, the particles are homogeneously distributed across the pipe as a result of turbulent mixing. In Case 4, with a bulk solid volume fraction of 0.195, the particles are aggregating in the pipe centre with a local volume fraction nearly twice as high as the bulk volume fraction. For Case 5 the aggregation is even more pronounced. The velocity fluctuations in the core are strongly damped (as observed from the uniform color in the core), indicating that the core is not turbulent anymore. Cases 6 and 6* are the cases with the highest bulk solid volume fraction. As in the other cases, particles have migrated to the core and seem to have formed a solid plug flowing at a nearly uniform velocity. Interestingly, for Case 6* with frictionless particles, the pipe wall is lined with patches of ordered particle structures. These ordered patches are not observed for Case 6 with frictional particles. In Case 6* also more particles are observed in the vicinity of the pipe wall.

\subsection{\label{Sect:2Dmaps}Time-averaged velocity and solid volume fraction maps}
\begin{figure}[ht]
    \centering
    
    \subfloat[]{
    \includegraphics[width=0.48\textwidth]{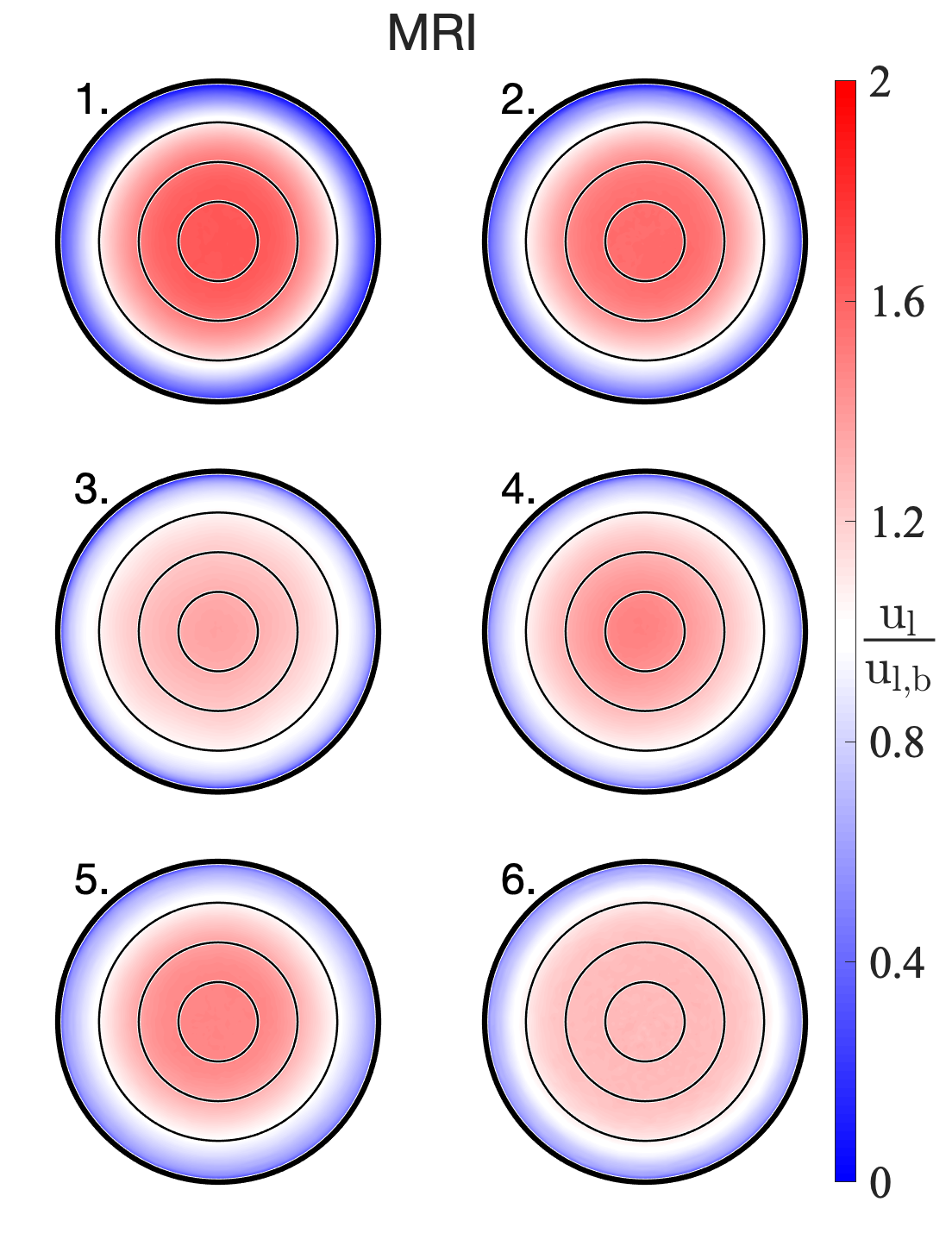}
    \label{fig:4a}}
    \hfill
    \subfloat[]{
    \includegraphics[width=0.48\textwidth]{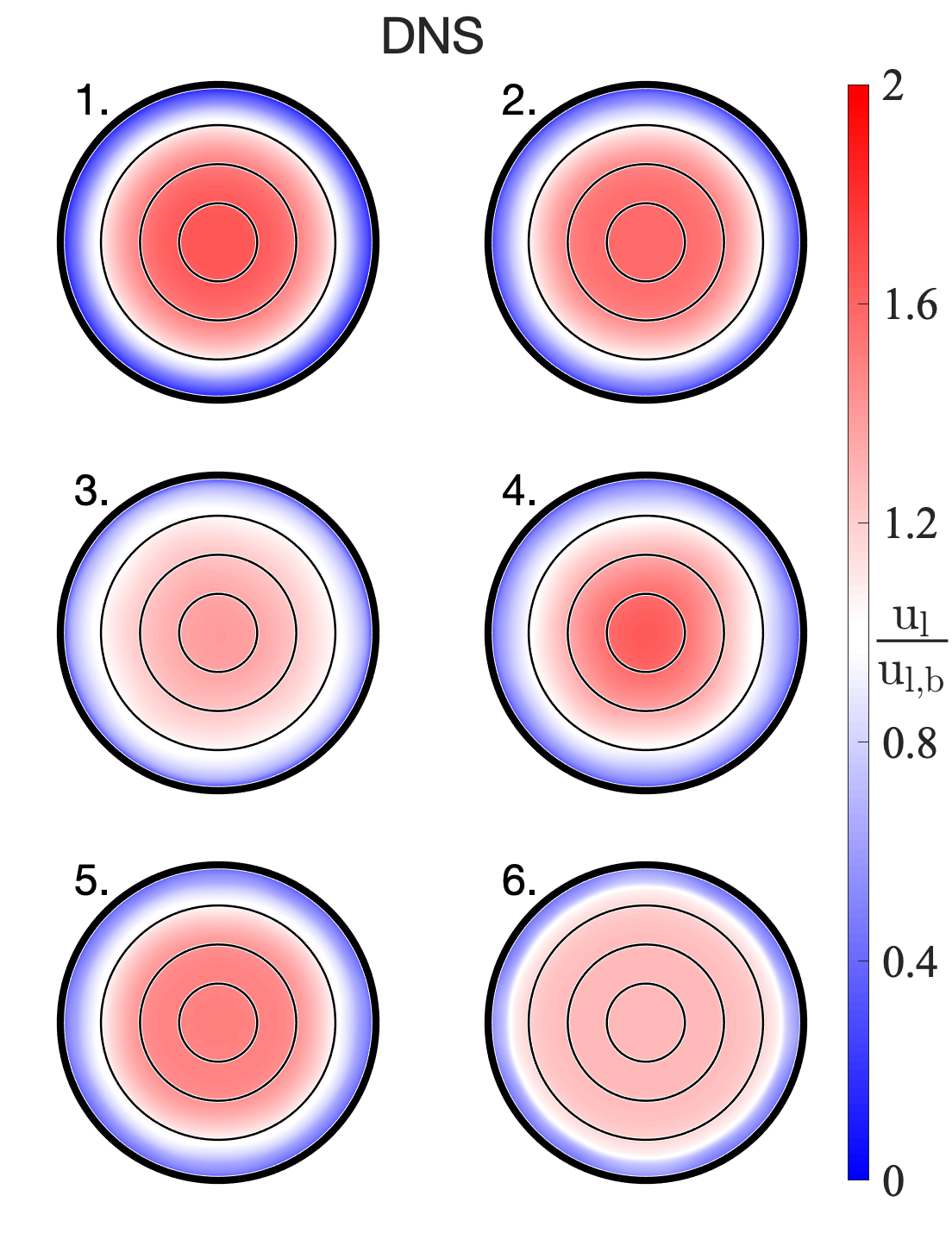}
    \label{fig:4b}}
    \caption{Time-averaged intrinsic liquid velocity distributions \protect\subref{fig:4a} based on the MRI experiments and \protect\subref{fig:4b} the numerical simulations. The solid rings are added to identify different radial locations, $r/R$ = 0.25, 0.5, and 0.75, in order to assess the axisymmetry of the different cases.}
    \label{fig:4}
\end{figure}

The time-averaged normalized intrinsic liquid velocity distributions ($u_l/u_{l,b}$) for the different cases for the experimental and numerical results are shown in Fig.~\ref{fig:4a} and \ref{fig:4b}, respectively. These figures clearly show the axisymmetry of the flow. As visual guide, black rings are superimposed at three radial locations ($r/R$ = 0.25, 0.5, and 0.75). These results show that there are no systematic structures or consistent deviations present in the measurements; these could have been caused by measurement artefacts or by a density mismatch.
From the color, the degree of blunting of the velocity profiles can be seen. In particular, Case 6 ($Re_s$ = 1732, $\phi_b$ = 0.466) exhibits a nearly uniform velocity distribution for $r/R<0.75$. From these visualisations it is evident that there is a good agreement between experiments and simulations, which will be quantified in Sect.~\ref{Sect:comparison}.

\begin{figure}[ht]
    \centering
    \subfloat[]{
    \includegraphics[width=0.48\textwidth]{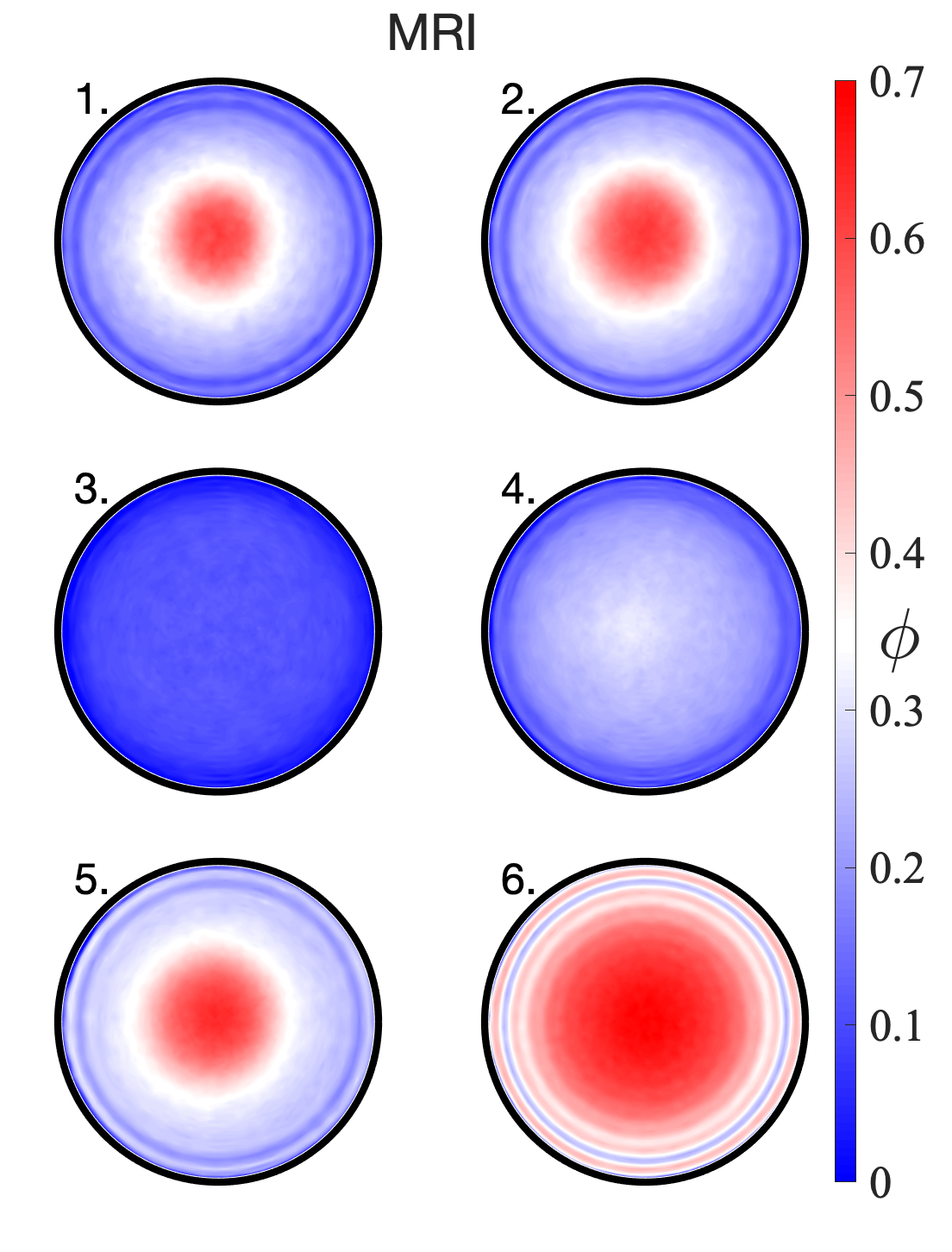}
    \label{fig:5a}}
    \hfill
    \subfloat[]{
    \includegraphics[width=0.48\textwidth]{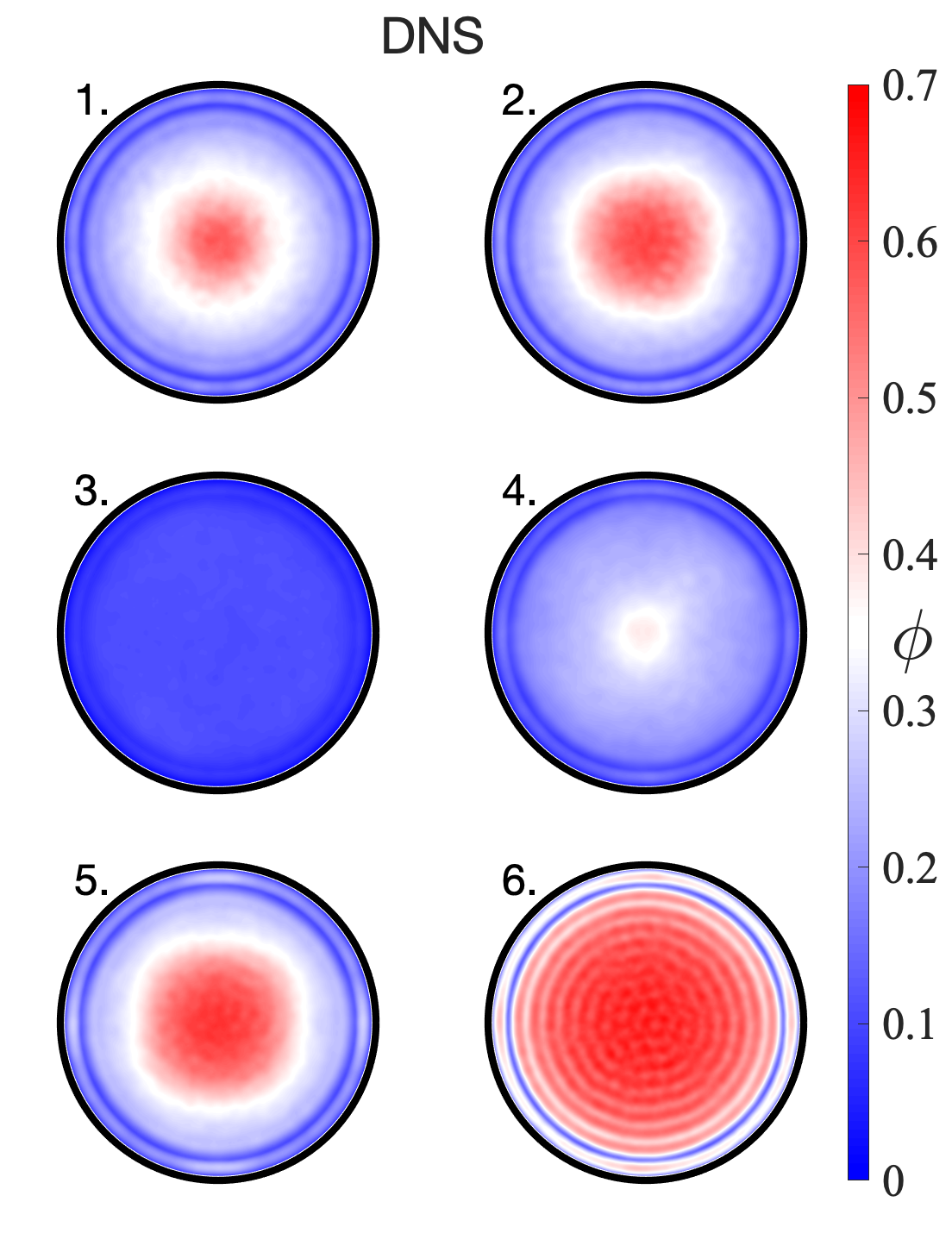}
    \label{fig:5b}}
    \caption{Time-averaged volume fraction distributions for the \protect\subref{fig:5a} experimental and \protect\subref{fig:5b} numerical results.}
    \label{fig:5}
\end{figure}

Fig.~\ref{fig:5}, presents the time-averaged solid volume fraction distributions for the experimental and simulation results. Again, it can be seen that there is a good agreement between the experimental and numerical results. Note that Case 1, 2, and 5 ($\phi_b$ = 0.25-0.31) are qualitatively similar, all exhibiting a core-peaking particle distribution despite different $Re_s$. The relatively uniform volume fraction distribution for Case 3 can be explained by the relatively low bulk solid volume fraction ($\phi_b$ = 0.089) in combination with a higher Reynolds number. The particles are homogeneously distributed due to turbulent mixing. Also, the effect of an increase in $\phi_b$ for approximately constant $Re_s$ in Cases 3-5 is clearly visible. This shows the change from a uniform to a core-peaking distribution for increasing $\phi_b$. Case 6 differs from the other cases, as ring-like structures are prominently visible here (even more pronounced in the DNS).

\subsection{\label{Sect:1Dprofiles}Time-averaged velocity and solid volume profiles}
\begin{figure}[!ht]
    \centering
    
    \subfloat[]{
    \includegraphics[width=0.45\textwidth]{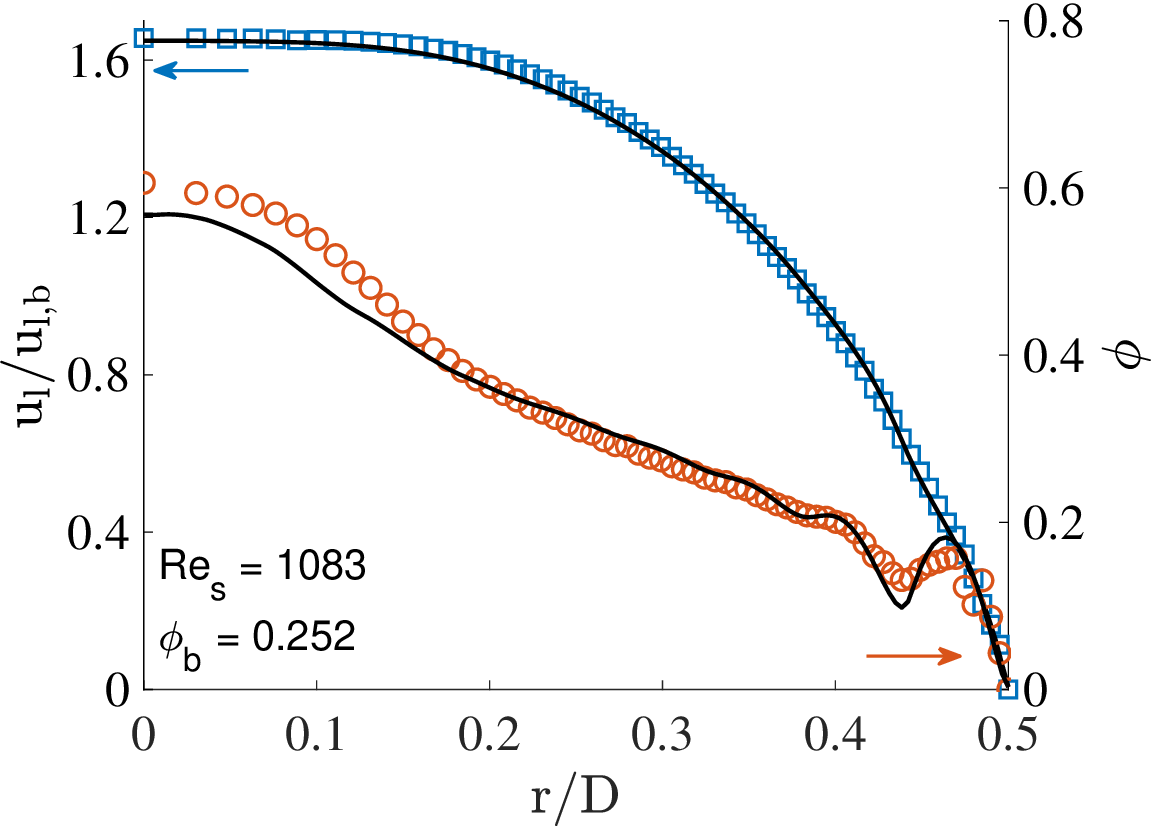}
    \label{fig:6a}}
    \hfill
    \subfloat[]{
    \includegraphics[width=0.45\textwidth]{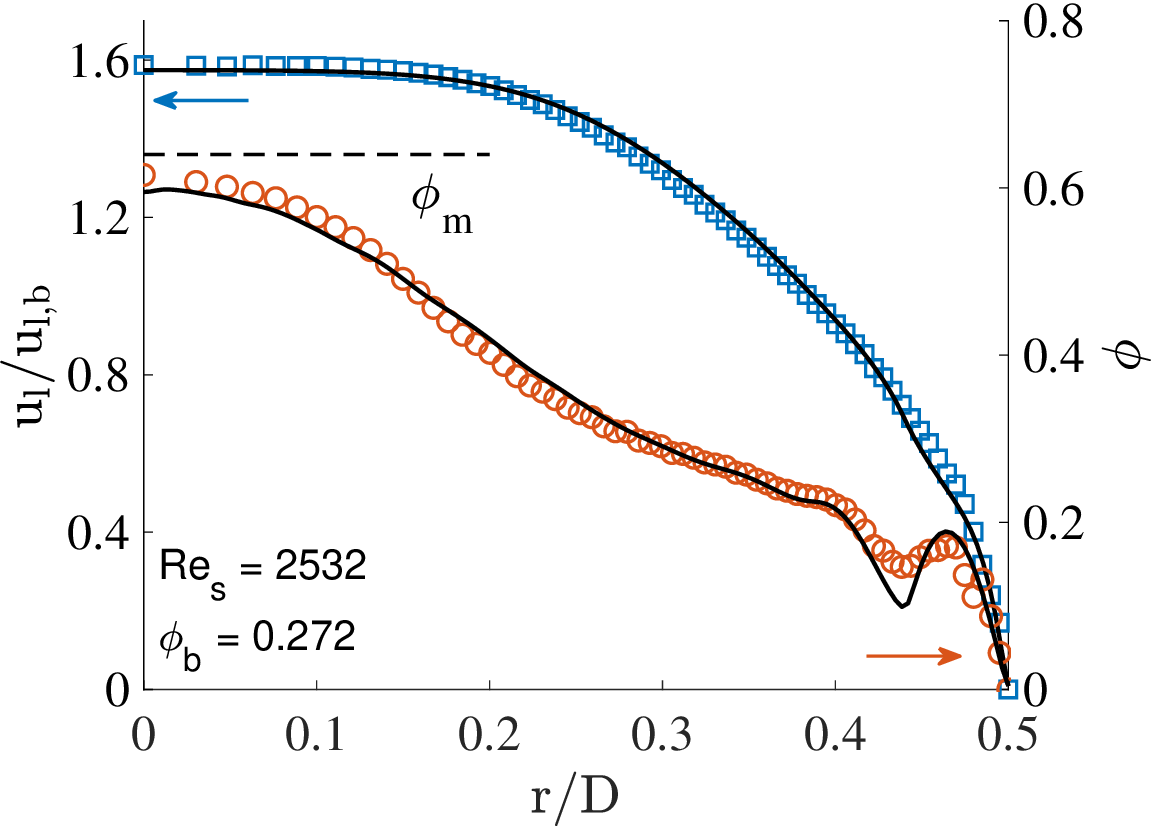}
    \label{fig:6b}}
    \hfill
    \subfloat[]{
    \includegraphics[width=0.45\textwidth]{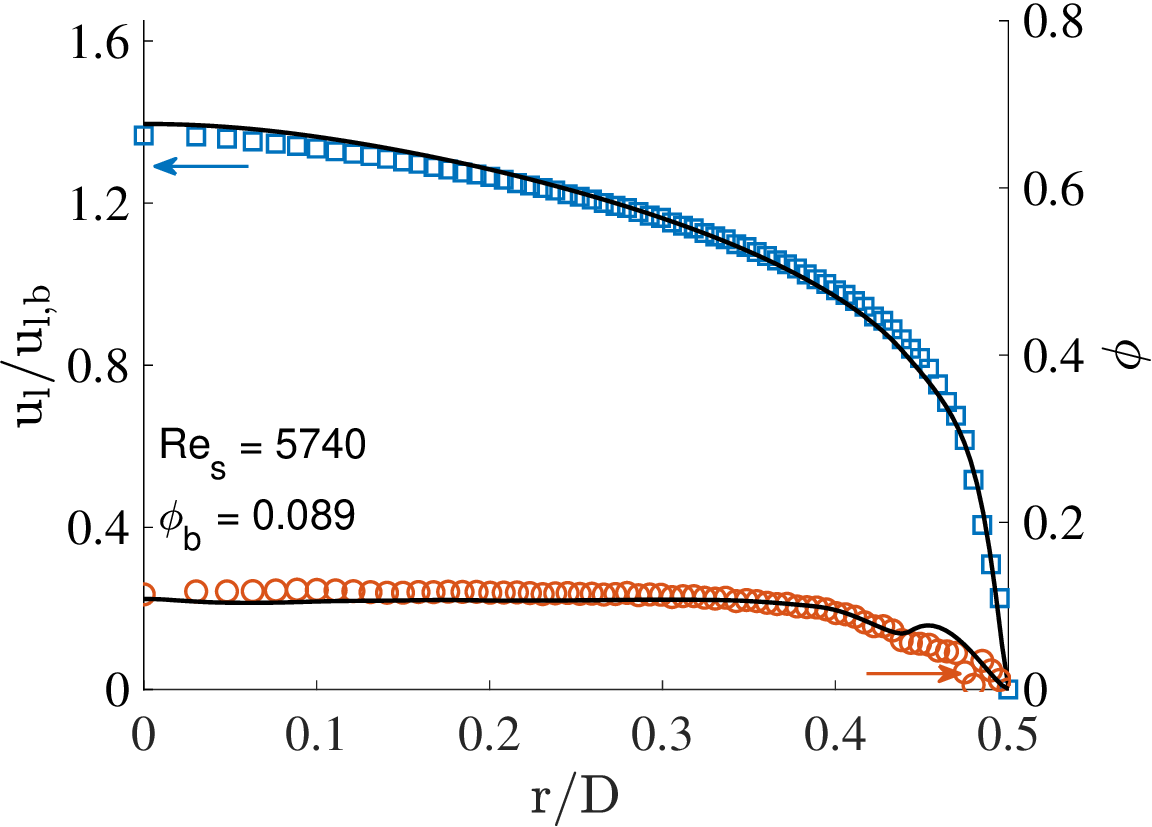}
    \label{fig:6c}}
    \hfill
    \subfloat[]{
    \includegraphics[width=0.45\textwidth]{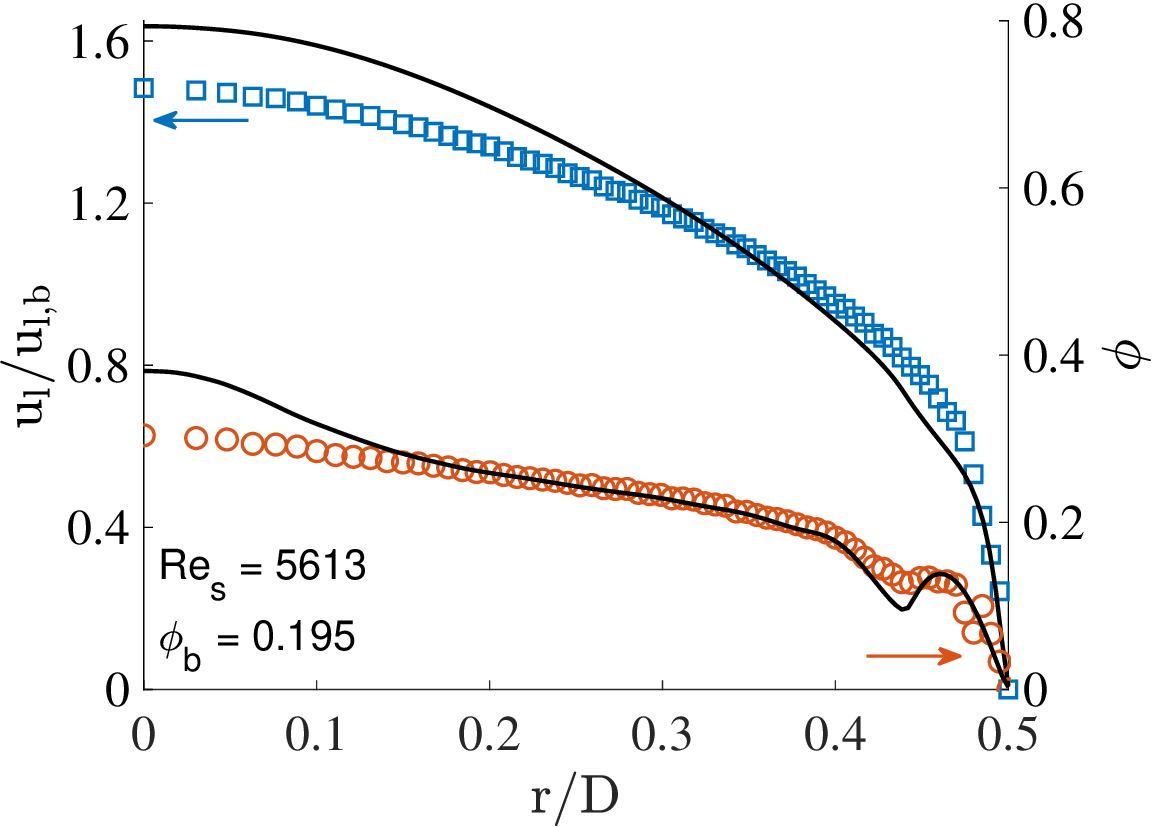}
    \label{fig:6d}}
    \hfill
    \subfloat[]{
    \includegraphics[width=0.45\textwidth]{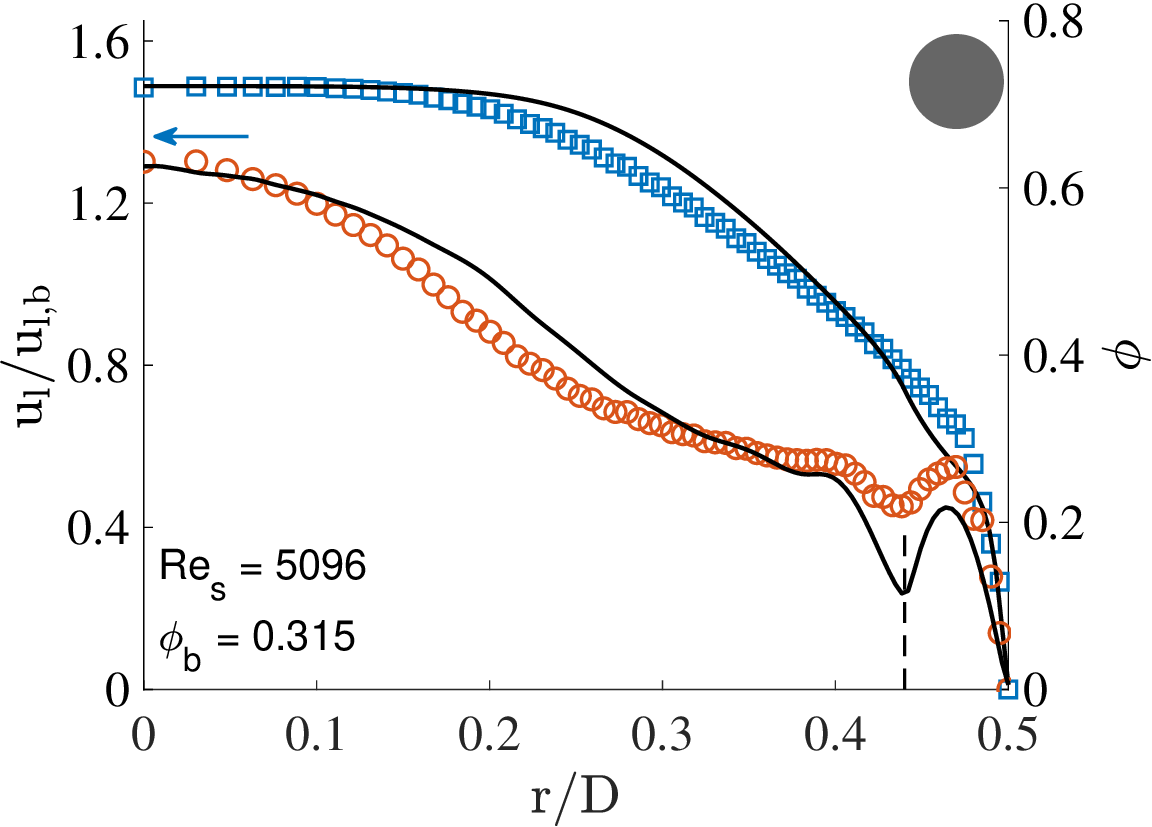}
    \label{fig:6e}}
    \hfill
    \subfloat[]{
    \includegraphics[width=0.45\textwidth]{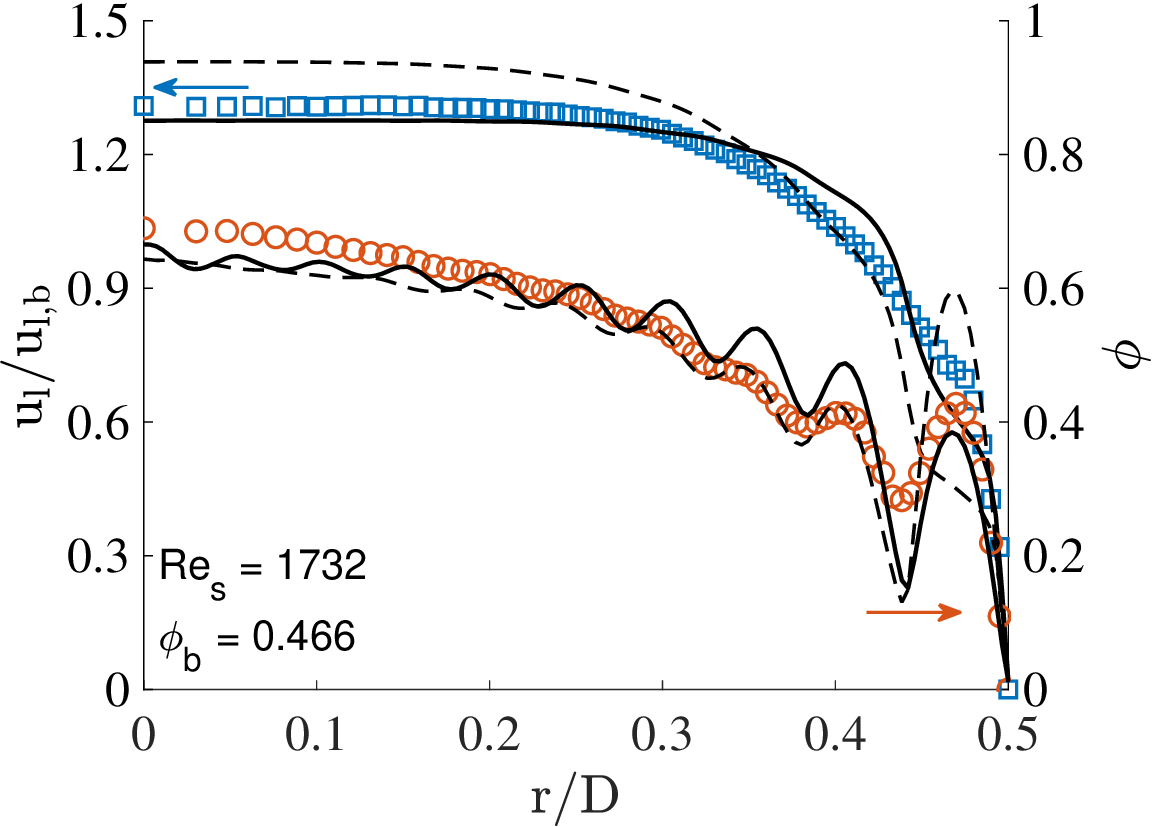}
    \label{fig:6f}}
    \caption{Normalised intrinsic liquid velocity and solid volume fraction profiles for \protect\subref{fig:6a} Case 1, \protect\subref{fig:6b} Case 2, \protect\subref{fig:6c} Case 3, \protect\subref{fig:6d} Case 4, \protect\subref{fig:6e} Case 5, and \protect\subref{fig:6f} Case 6. The experimental velocity (square markers) and solid volume fraction (round markers) results are corresponding to the left-hand-side and right-hand-side y-axis, respectively. The DNS results are represented by the solid curves. The dashed curve in \protect\subref{fig:6f} corresponds to the additional frictionless simulation, Case 6$^*$. The flow conditions for the experiments are listed in the figures. Note that the DNS settings may slightly deviate from this (cf. Table \ref{table:2} and \ref{table:2b}).}
    \label{fig:6}
\end{figure}

A quantitative representation of the time-averaged velocity (Fig.~\ref{fig:4}) and solid volume fraction (Fig.~\ref{fig:5}) distributions is presented in Fig.~\ref{fig:6}, where these distributions are azimuthally averaged. The normalised velocity and solid volume fraction profiles for Case 1 ($Re_s$ = 1083) and Case 2 ($Re_s$ = 2532) are shown in Fig.~\ref{fig:6a} and \ref{fig:6b}, respectively. The experimental results are represented by the markers and the DNS results are shown by the solid lines. The velocity profiles correspond to the left-hand-side ordinate axis, whereas the right-hand-side ordinate axis corresponds to the solid volume fraction profile.
Note that the velocity and solid volume fraction profiles of both cases are very similar, despite the difference in Reynolds number. The flow behavior is apparently governed by the particle volume fraction, rather than $Re_s$ (this will later on be confirmed by Case 5). There is a very good agreement between the experimental and numerical velocity profiles. Also, the trends in the solid volume fraction distribution are in good quantitative agreement. In the near-wall region and the pipe centre there are slight deviations. This will be further elaborated upon later (see, Sect.~\ref{Sect:comparison}) as several reasons are likely responsible for this deviation.
For both cases a strong concentration gradient can be observed. The solid volume fraction at the pipe centre is more than twice the bulk volume fraction, resulting from shear-induced and inertia-driven migration. Furthermore, the `wiggle' in the near-wall region can be associated with the presence of a particle wall layer. Note that the minimum in $\phi_b$ occurs at $r/D \approx$ 0.44, which corresponds to one particle diameter from the wall: $r/D$ = $R/D$ - $d/D$ = 0.5-0.058 = 0.442. This location is indicated with a vertical dashed line in Fig.~\ref{fig:6e}. Also, in this figure a particle is added in the top right corner. The wall constraints the particles, causing them to order in a ring-like structure (as was also pointed out before by \citet{hampton1997migration}).
Indirectly, the results also confirm that the flows are fully developed. This can be inferred from the agreement between the experimental and numerical results and the fact that the numerical results are fully converged. This confirms that in the experiments a pipe length of 132$D$ was sufficient for the suspension to reach an equilibrium.
Furthermore, it can be seen that the high volume fraction at the pipe centre is responsible for the blunting of the velocity profile due to the locally high solid stress. The particle stacking in the pipe centre is close to the random close packing fraction ($\phi_m \approx$ 0.64), and thereby highly limiting shear due to the very high local suspension viscosity. As a visual guide this maximum packing fraction is indicated with a horizontal dashed line in Fig.~\ref{fig:6b}. To illustrate this, for $\phi_b >$ 0.4, the average nearest-neighbor distance between uniformly distributed finite particles is less than one tenth the particle diameter \cite{bansal1972average}. This highly limits the possibility for the particles to get advected with respect to each other due to a velocity gradient. The velocity gradient rapidly increases for decreasing local volume fraction, as can be observed for the lower local volume fractions for higher $r/D$. \\

The comparison for the following three cases (Case 3 -- 5) are presented in Fig.~\ref{fig:6c}-\ref{fig:6e}. For these cases the bulk solid volume fraction is increasing for nearly constant $Re_s$. For increasing $\phi_b$, the volume fraction distribution gradually changes from a homogeneous distribution (Case 3) via a mild gradient in the solid volume fraction distribution (Case 4) to a distinct core-peaking distribution (Case 5). In general, the experimentally and numerically obtained results are in qualitative agreement, as the trends are well captured. The deviation between both data sets for Case 4 can be explained by the transitional nature for these specific conditions, and will be elaborated further upon later. Note that the radial volume fraction distribution of Case 5 is very similar to the distributions seen before in Case 1 and 2. This suggests that beyond a certain bulk volume fraction the solid volume fraction distribution and the velocity distribution are mainly governed by $\phi_b$ rather than $Re_s$. \\

The comparison for the final case, also being the most extreme case studied, is shown in Fig.~\ref{fig:6f}. This case can be considered to be on the edge of the parameter space, as for the high solid volume fraction, $\phi_b$ = 0.466, the experimental setup - specifically the inlet chamber - was close to jamming. Also for this case, the trends in the velocity and solid volume fraction distributions for the experiments and simulations are in qualitative agreement. Because of the relatively high solid volume fraction, inter-particle friction might be particularly important for this specific case. In order to study the influence of this frictional contribution, a DNS study of a frictionless case (i.e., Case 6$^*$) is performed. The result of this case is represented by the dashed line. Although there is still a discrepancy between the frictionless case and the experimental results, the `wiggles' in the solid volume fraction profile are less pronounced and in better agreement with the experimental results. As pointed out in Sect.~\ref{Sect:3DFields} patches of ordered particle structures are observed in the frictionless case (Case 6$^*$). As a result of this, more particles are observed in the vicinity of the pipe wall. This is also visible in the relatively high concentration peak near the wall for this case. The `wiggles' in the concentration profile from the DNS of Case 6 indicate that the effect of particle layering is not only felt at the wall but even in the core. Furthermore, the difference in the peak concentration for the core between the experiments and the DNS suggests that the maximum flowable packing fraction is somewhat lower in the DNS than in the experiments.

\subsection{\label{Sect:comparison}Comparison between MRI and DNS}
\begin{figure}[ht]
    \centering
    
    \subfloat[]{
    \includegraphics[width=0.48\textwidth]{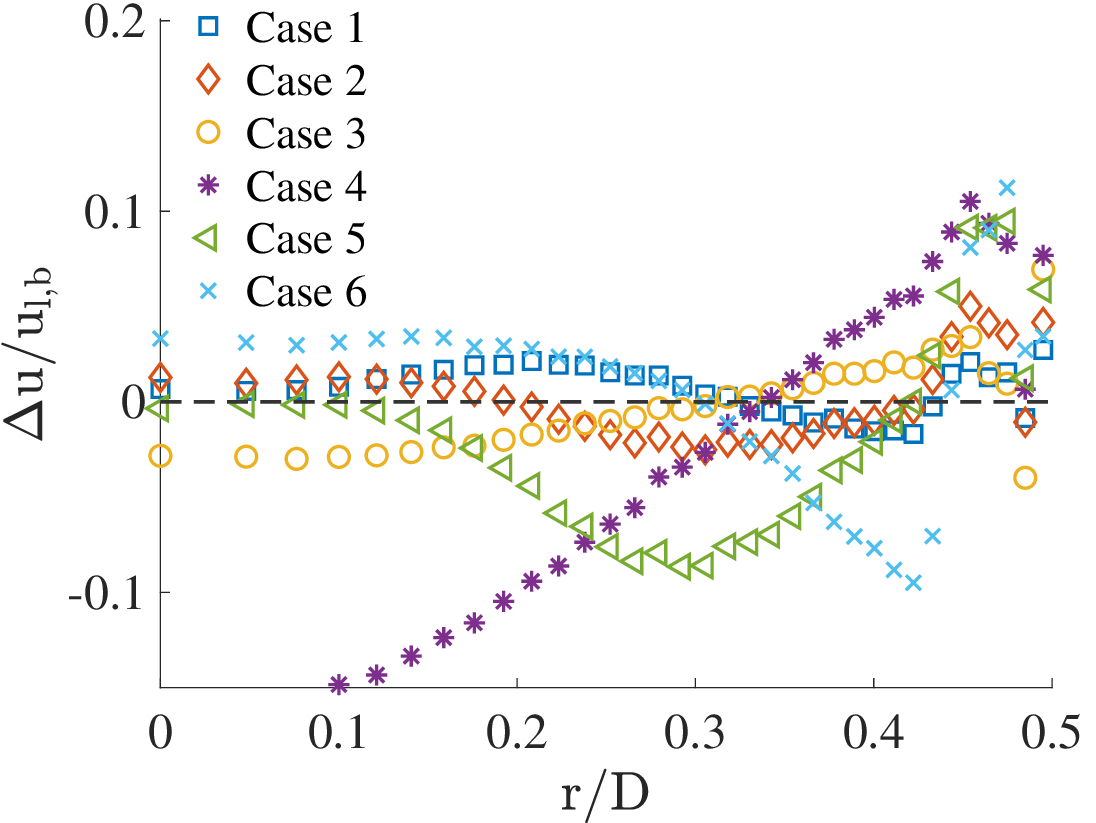}
    \label{fig:7a}}
    \hfill
    \subfloat[]{
    \includegraphics[width=0.48\textwidth]{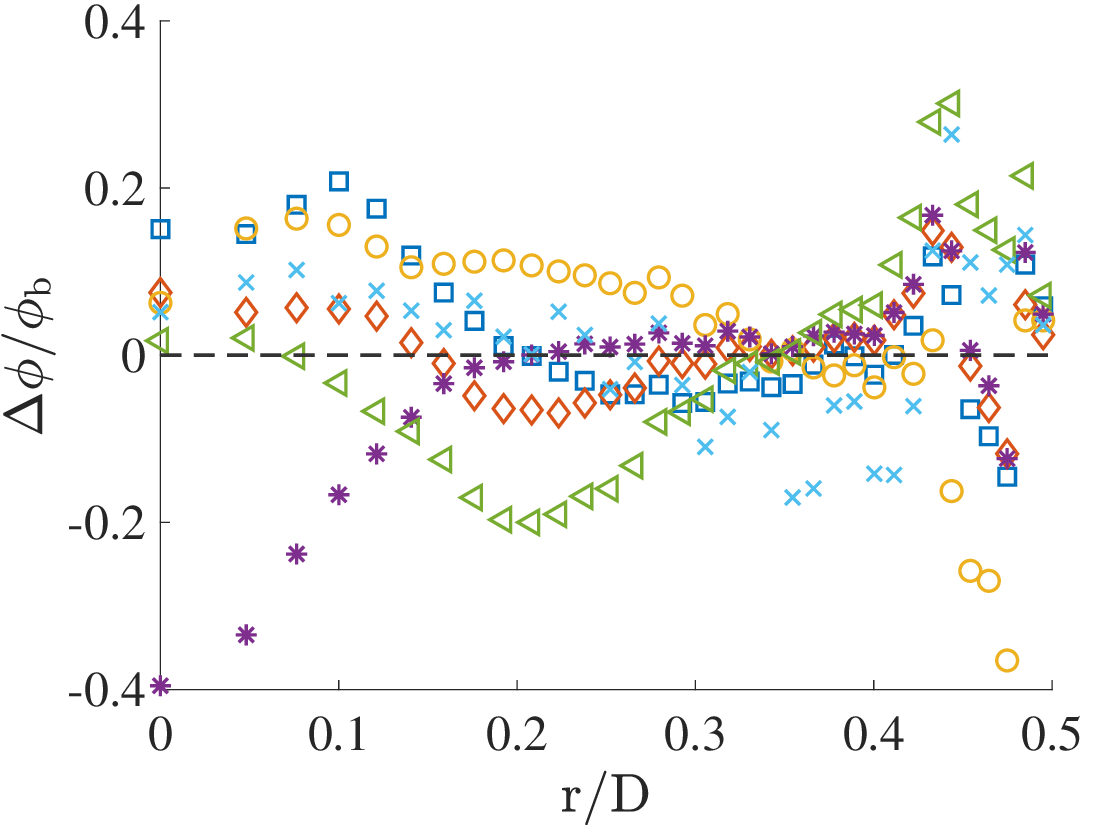}
    \label{fig:7b}}
    \caption{Normalised deviation between the experimental and numerical results for the \protect\subref{fig:7a} velocity and \protect\subref{fig:7b} solid volume fraction for the studied cases (indicated in the legend). For each case every second marker is shown to improve readability.}
    \label{fig:7}
\end{figure}

In general, there is a good agreement between the experimental and numerical results. The deviation between the experimental and numerical results are quantified using the RMS value of the difference between both results. The difference between the experimental and numerical results for the velocity and solid volume fraction profiles are shown in Fig.~\ref{fig:7a} and \ref{fig:7b}, respectively. The corresponding RMS values are listed in Table \ref{table:3}. The maximum normalised error between the velocity profiles is found to be 8.0\% for Case 4. For the other cases the error is less than 5\%, with an error of only 1.2\% for Case 1.

\begin{table}[htb]
	\centering
	\caption{Normalised root mean square error of the difference between experimental and numerical results for the liquid velocity and solid volume fraction.}
	\begin{tabular}[t]{p{2cm} p{4cm} p{4cm} }
	\toprule
	\textbf{Case}   &\textbf{$\sqrt{\overline{\Delta u(r)^2}}/u_{l,b}~[\%]$}     &\textbf{$\sqrt{\overline{\Delta \phi(r)^2}}/\phi_b~[\%]$}  \\
	\midrule
	1               & 1.2                & 8.4\\
	2               & 2.0                & 5.8\\
	3               & 2.1                & 12.6\\  
    4               & 8.3                & 11.2\\ 
	5               & 5.1                & 13.4\\  
	6               & 5.0                & 9.7\\ 
	\bottomrule
    \end{tabular}
	\label{table:3}
\end{table}

The discrepancies between the experimental and numerical can be explained by various reasons. In short we have identified four different causes which likely contribute to the differences observed.
In the first place, there is a dissimilarity in particle size-distribution between the experimental and numerical approach, which are polydisperse and monodisperse systems, respectively. More specifically, this difference will have two effects: (i) it will affect the spatio-temporal solid volume fraction distribution, and (ii) it changes the maximum random close packing fraction, as this is a function of the particle size distribution (more in particular the polydispersity and skewness) \cite{desmond2014influence}. The result of (i) is in particular visible in dense systems and in the near-wall region (see, e.g., Case 6). For monodisperse suspensions radial variations of the average volume fraction are expected to be more pronounced than for polydisperse suspensions because of the more random ordering of particles with various sizes. For (ii), the maximum random close packing fraction of the polydisperse system is determined to increase by 1.3\% with respect to the monodisperse system \cite{desmond2014influence}. Note, however, that the correlation from that study is based on `dry' particles being packed in the absence of driving forces. This is in contrast to the current study, where `wetted' particles are packed under the influence of a driving force (shear-induced migration). Additionally, the suspension is also flowing (dynamic vs. static), which might affect the particle packing, in particular when transient effects are concerned. However, these differences are expected to be of minor influence on the maximum random close packing fraction, also in view of experimentally determined maximum solid volume fractions (see, e.g., Fig.~\ref{fig:6f}). The change of the maximum random close packing will start to play a role for cases where the local volume fraction is approaching this maximum random packing fraction (e.g., Cases 1, 2, 5 and 6). This partially explains the slightly higher solid volume fractions for the experiments at the pipe centre for these cases. In turn, this higher volume fraction at the pipe centre needs to be compensated for at a different radial location, because the experimental and numerical cases are compared for the same bulk solid volume fraction.
Secondly, the particle roughness effects on lubrication, and interparticle and particle-wall frictional collisions in stick and slip regimes is a potential cause for differences. In the DNS the Coulomb coefficient of sliding friction has been set to 0.39 based on in-house experiments by \citet{shajahan2023} as mentioned earlier. The frictionless simulation of Case 6$^*$, with a Coulomb coefficient of sliding friction of 0, reveals that this coefficient has important implications for the modeling and dynamics of these dense suspensions. 
In the third place, the uncertainty in the experimental suspension flow parameters (e.g., $\phi_b$, $u_{l,b}$, etc.) can be identified as a source of error. As the DNS is based on these experimental parameters, this might result in a mismatch between experimental and numerical results. This mismatch is in the order of the experimental uncertainty as described in the experimental setup section. 
A final explanation might be the numerical resolution and limited streamwise domain length (with periodic boundary conditions) in DNS: this could be further analysed by means of streamwise autocorrelation functions and energy spectra of the streamwise particle, fluid and/or mixture velocities.

Besides the above-mentioned reasons for discrepancies between the experimental and simulation results, there is one additional reason for the discrepancy for Case 4. This deviation is likely explained by the transitional nature for these specific flow conditions. The solid volume fraction distribution for this case is in between the homogeneous ($\phi_b$ = 0.089) and core-peaking ($\phi_b$ = 0.315) case, as seen in Fig.~\ref{fig:6}. From the experimental results for Case 4 ($\phi_b$ = 0.195) there is already some evidence of the core-peaking process, as a slight concentration gradient is already present. The increase in the local volume fraction at the pipe center from $\phi_b$ = 0.304 (Case 4) to $\phi_b$ = 0.631 (Case 5) occurs for an increase of 12\% in the bulk solid volume fraction. This supports our suggestion that Case 4 is a transitional case, and which is thus an especially sensitive case for comparison.\\ 

While the MRI and DNS results for the radial concentration and liquid velocity profile are in good agreement with each other, it remains puzzling why large differences are found for the Darcy-Weisbach friction factor. This holds in particular for Cases 5, 6 and 6*, where $\phi_b$ is high. This suggests that the overprediction of the friction factor in the DNS originates from a much stronger solid stress and hence higher suspension viscosity as currently predicted from Eilers' correlation with $\phi_m = 0.64$. 
Indeed, the solid volume fraction profile in Figure~\ref{fig:6f} suggests that the maximum flowable packing fraction is somewhat less in the DNS than in the experiments, which is consistent with a significantly higher suspension viscosity in the DNS for the most dense cases. Furthermore, the difference in friction factor between Cases 6 and 6* indicates a significant influence of the Coulomb coefficient of sliding friction, $\mu_c$, on the solid stress. We are currently exploring the effect of $\mu_c$ on other cases. Finally, from a preliminary study it appears that the influence of the other collision parameters on the flow dynamics, i.e., the normal and tangential coefficients of restitution, is small.

\subsection{\label{Sect:Model}Characteristic cases and parameter space}
\begin{figure}[ht]
    \centering
    
    \subfloat[]{
    \includegraphics[width=0.48\textwidth]{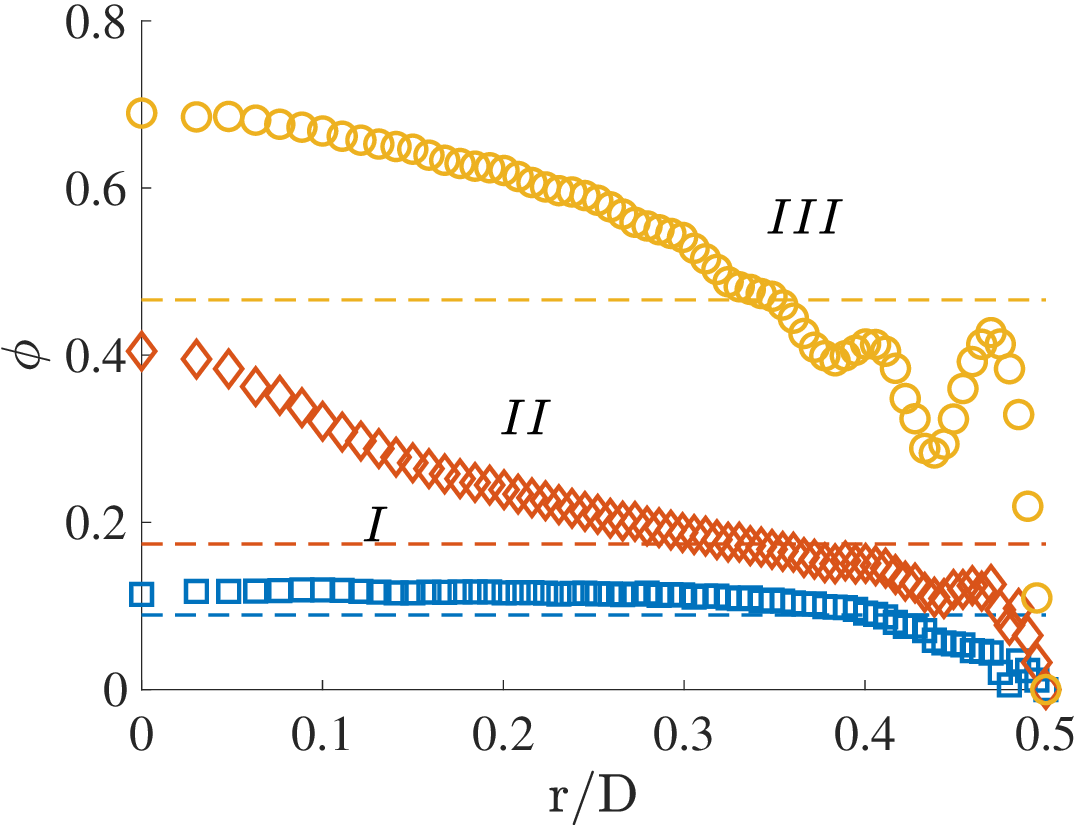}
    \label{fig:8a}}
    \hfill
    \subfloat[]{
    \includegraphics[width=0.48\textwidth]{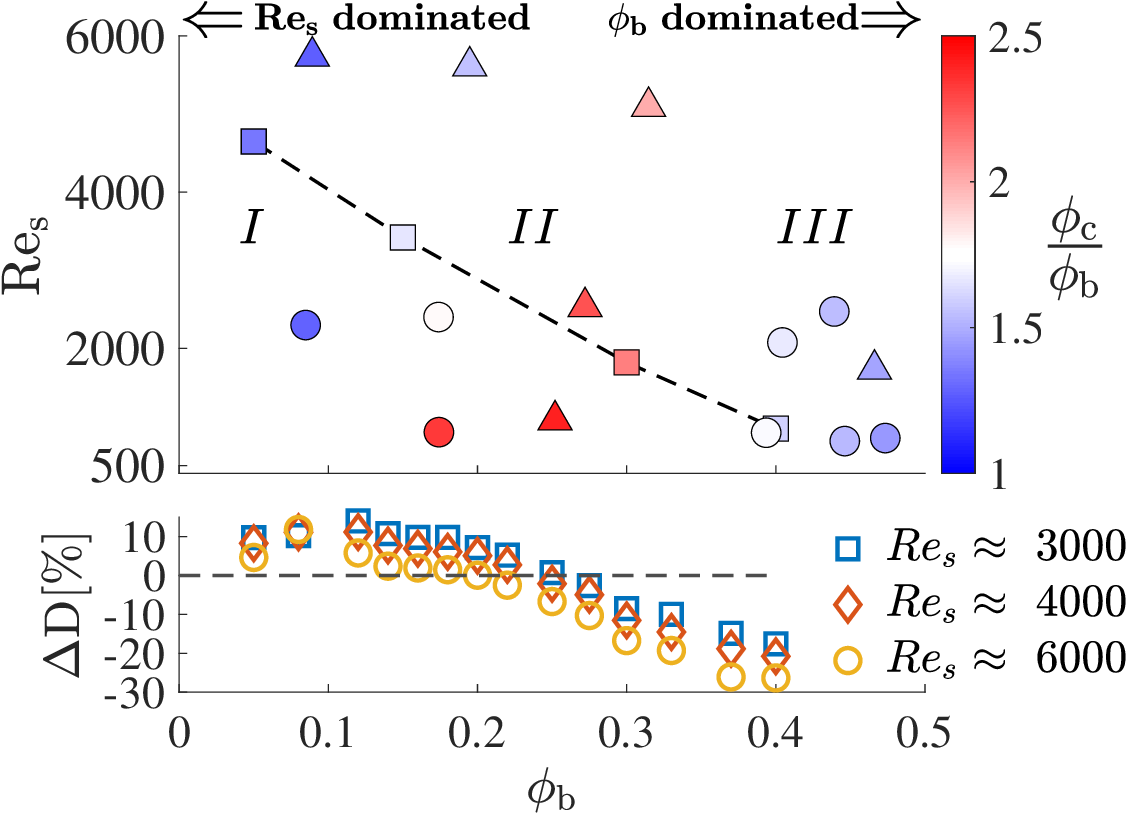}
    \label{fig:8b}}
    \caption{Characteristic solid volume fraction distributions for three characteristic cases \protect\subref{fig:8a}. The horizontal dashed lines represent the corresponding average solid volume fraction. The regime map corresponding to these cases is shown in the top panel in \protect\subref{fig:8b}. The triangular markers are the six cases studied above. The round markers in the regime map are additional experimental results (exact conditions are listed in the Appendix). The square markers, connected with a dashed line, are numerical results for a similar pipe flow study ($d/D$ = 1/15) from \citet{ardekani2018numerical}. The regimes are indicated with roman numerals, corresponding to the profiles in \protect\subref{fig:8a}. The bottom panel of \protect\subref{fig:8b} shows the drag change with respect to the single phase drag for increasing $\phi_b$.}
    \label{fig:8}
\end{figure}

The six cases compared above are selected from a larger experimental data set. Based on this data set and additional numerical data from literature \cite{ardekani2018numerical}, we can distinguish between three different characteristic cases. The solid volume fraction profiles of these cases, indicated with \MakeUppercase{\romannumeral 1}, \MakeUppercase{\romannumeral 2}, and \MakeUppercase{\romannumeral 3}, are visualised in Fig.~\ref{fig:8a}. The dashed lines are the corresponding bulk solid volume fractions, added for reference. For relatively \textit{low} solid volume fractions and \textit{higher} $Re_s$ a nearly uniform particle distribution is observed (Case 3). Here turbulence is dominant and causing a relatively homogeneous particle distribution (case \MakeUppercase{\romannumeral 1}). For \textit{moderate} $\phi_b$ the particles are found to aggregate to the pipe centre, forming a solid particle core. This behavior is observed for the Cases 1, 2, and 5 (case \MakeUppercase{\romannumeral 2}). Eventually, for \textit{high} $\phi_b$, the particle core is expanding in the direction of the wall, as the maximum packing fraction at the pipe centre is reached (see, e.g., Cases 6 and 6*); this will be referred to as case \MakeUppercase{\romannumeral 3}.

These characteristic cases are summarized in a $\phi_b$ vs. $Re_s$ parameter space, shown in the top panel of Fig.~\ref{fig:8b}. The triangular markers correspond to the conditions of the six cases studied above. The round markers are additional experimental results, the exact flow conditions of these experiments are listed in Appendix~\ref{A:1}. Note that in total 14 multiphase cases are studied experimentally, from which six cases are compared to DNS. Also, results from a similar numerical study (slightly different $d/D$ of 1/15) by \citet{ardekani2018numerical} are added in the regime map, shown as square markers connected by a dashed line. Note that these simulations are all performed at constant $Re$ = 5300. For the comparison in Fig.~\ref{fig:8b} a viscosity correction is applied using Eilers' model \cite{eilers1941viskositat}. For the maximum packing fraction, $\phi_m$ = 0.64 is used, as this is the maximum packing fraction at the pipe centre in their study. This explains their decreasing $Re_s$ for increasing $\phi_b$. The marker color represents the ratio between the solid volume fraction at the pipe center, $\phi_c$, and the bulk solid volume fraction, $\phi_b$. For uniformly distributed systems this ratio approximates 1, while for core-peaking systems this value will exceed 2 (see, e.g., case \MakeUppercase{\romannumeral 2} in Fig.~\ref{fig:8a}). For higher $\phi_b$, the amplitude of this ratio decreases again, as the maximum packing fraction at the core is reached and the core is expanding in the direction of the pipe wall. This can indeed be observed for increasing $\phi_b$ (see, e.g., Fig.~\ref{fig:6}). Note the consistency between the marker colors of the present study and the marker colors of the study by \citet{ardekani2018numerical}. This parameter allows for a general classification of the different regimes. These regimes, indicated with the Roman numerals \MakeUppercase{\romannumeral 1}, \MakeUppercase{\romannumeral 2}, and \MakeUppercase{\romannumeral 3}, are corresponding to the characteristic cases in Fig.~\ref{fig:8a}. Based on the analysis of Fig.~\ref{fig:6}, the flow is $Re_s$ dominated for low $\phi_b$, while for high $\phi_b$ the flow is dominated by $\phi_b$ itself. See e.g., the comparison between Case 1, 2, and 5. Similar observations are presented in the regime-map for channel flow in a study by \citet{lashgari2014laminar}.

The bottom panel in Fig.~\ref{fig:8b} presents the drag change, $\Delta D$, for increasing $\phi_b$ for various $Re_s$. $\Delta D$ is defined as the drag change with respect to the drag for the same $Re_s$ single-phase flow, obtained using Blasius' correlation. Note that these results are obtained from a separate series of pressure drop experiments, which explains the additional data points. The trend in the drag change confirms earlier observations by \citet{hogendoorn2018particle} and \citet{agrawal2019transition}. Interestingly, the drag change can be related to the characteristic solid volume fraction distributions from Fig.~\ref{fig:8a}. For nearly uniform systems (region \MakeUppercase{\romannumeral 1}), a drag \textit{increase} is observed. Note that this drag increase is not simply due to the enhanced effective suspension viscosity, which is taken into account. For the most extreme case this drag increase is 15\% with respect to single-phase flow. For $\phi_b \gtrapprox$ 0.12 (region \MakeUppercase{\romannumeral 2}) the drag is found to \textit{decrease} again. For higher $\phi_b$ (region \MakeUppercase{\romannumeral 3}), the drag \textit{decreases} further, even significantly below the single-phase case (i.e., up to 25\%). 

We speculate that the increasing and decreasing drag curves can be explained by a balance between two competing mechanisms: for \textit{low} $\phi_b$, additional friction is introduced by the particle layer lining the pipe wall that acts as a rough and porous wall layer \cite{costa2016universal,picano2015turbulent}, while for \textit{higher} $\phi_b$  
solid volume fraction gradients are formed, which cause a strong non-uniform effective viscosity in the radial direction for high $\phi_b$. The relatively low $\phi_b$ in the near-wall region compared to the core acts as a lubrication layer between the pipe wall and the dense particle core. This is similar to core-annular flow, where drag reduction is accomplished by a low viscosity lubrication layer \cite{oliemans1986core}. Note that particle fluctuations, inherently present in these flows \cite{hogendoorn2022onset}, cause mixing and thus may affect the particle distribution. This needs to be accounted for in a successful theoretical model.

Interestingly, when for higher $\phi_b$ the effective viscosity of the near-wall particle layer (i.e., 0.44 $< r/D <$ 0.5) is used to determine a new $Re_s'$, this $Re_s'$ is higher and the corresponding friction factor (determined using Blasius' correlation) agrees well with the measured friction factor. To illustrate, for Case 5 ($Re_s$ = 5096, $\phi_b$ = 0.315) with an average particle solid volume fraction at the wall of $\phi_w$ = 0.196, $Re_s'$ = 8340. This results in a drag difference in the order of 1\% with respect to Blasius' correlation, instead of the original 14\% shown in Fig.~\ref{fig:8b}. Note, however, that this is only valid in the region where $\phi_b$ is dominant. This shows that direct insight in the velocity and solid volume fraction profiles is required in order to understand the change in drag.

The data presented in this study provides a framework for a general regime classification. However, the exact regime boundaries are deliberately excluded in this study as additional data are required for a detailed definition of these boundaries. Various metrics can serve as an input for these boundaries. The change from region \MakeUppercase{\romannumeral 1} to region \MakeUppercase{\romannumeral 2} can be set at the location of the maximum drag, here the two competing mechanisms are balanced. The location where the maximum drag occurs is $Re_s$ dependent as can be seen from the bottom panel of Fig.~\ref{fig:8b}. For the boundary between region \MakeUppercase{\romannumeral 2} and \MakeUppercase{\romannumeral 3}, another metric can be used, for instance when the ratio $\phi_c/\phi_b$ drops below a certain threshold. This is an indication that the maximum packing fraction at the pipe centre is reached, and the particle core is expanding in the direction of the pipe wall.

\section{\label{Sect:conclusion}Conclusion}
Time-averaged intrinsic liquid velocity and solid volume fraction distributions of various suspension flows are experimentally and numerically studied. In total six cases are compared, where each case is characterized by a unique combination of $Re_s$ and $\phi_b$ for a fixed $d/D$. This allows us to study the effect of increasing $Re_s$ for approximately constant $\phi_b$, or increasing $\phi_b$ for approximately constant $Re_s$. Generally, a good agreement between the MRI and DNS results is found, which provides a solid basis for a further in-depth analysis. In particular, the results confirm earlier qualitative experimental, numerical and theoretical observations in literature, e.g., \cite{phillips1992constitutive,hampton1997migration,ardekani2018numerical,lashgari2016channel}.\\

Overall good quantitative agreement between the MRI and DNS results is found, with RMS-errors as low as 1.2\% and 8.4\% for the velocity and solid volume fraction profiles, respectively. The discrepancies between the experimental and numerical results is attributed to various reasons, including the difference in particle size distribution and the related maximum random packing fraction, slight deviations in particle roughness effects on lubrication and interparticle friction, uncertainty in experimental suspension flow parameters, and the numerical resolution. The contribution of each component might be more or less dominantly present in the flow results obtained, also depending on the flow regime.\\

Based on the compared cases and additional experimental results, a $\phi_b$ vs. $Re_s$ regime map is introduced. Three different regimes are identified. For \textit{low} bulk solid volume fractions nearly uniformly distributed systems are observed. Here turbulence is responsible for the mixing of the suspension. For \textit{moderate} volume fractions shear-induced migration is observed, and the particles accumulate at the pipe centre. For \textit{high} volume fractions the maximum packing fraction at the pipe centre is reached and the core is expanding in the direction of the pipe wall. These time-averaged solid volume fraction profiles explain the change in drag for increasing $\phi_b$. Initially a drag increase with respect to single-phase flows is observed, which is found to decrease for higher $\phi_b$. For $\phi_b$ = 0.4 a drag decrease of 25\% is found. This drag increase and decrease is likely explained by a balance between two competing mechanisms. We speculate that for low $\phi_b$, additional friction is introduced by the particle layer lining the pipe wall which acts as a rough and porous wall layer, while for higher $\phi_b$ solid volume fraction gradients are formed, which cause a strong non-uniform effective viscosity in the radial direction for high $\phi_b$. The near-wall region with relatively low $\phi_b$ acts as a lubrication layer between the pipe wall and the dense particle core, resulting in a more efficient transportation of the suspension.\\


For future work, various research directions were formulated in this study. One open question is how particle-laden flows transitions from a uniformly distributed system to a core-peaking distribution, and the exact role of the diameter ratio, $d/D$. MRI-based experiments will shed light on the exact nature of this change in flow regime. Additionally, higher-order statistics will shed further light on the dynamics underlying the different suspension flow regimes. Therefore, in conjunction with DNS, MRI-based Reynolds-stress measurements in particle-laden flows are currently pursued.\\

The authors confirm contribution to the paper as follows:
study conception and design: WH, CP;
data generation and processing: DF, MB, WH, WPB;
analysis and interpretation of results: WH, CP, WPB;
draft manuscript preparation: WH, WPB.
All authors reviewed the results and approved the final version of the manuscript.\\

Underlying data are deposited in 4TU.ResearchData \cite{underlying_data}.

\begin{acknowledgments}
We would like to thank Mr S. van Baal (SynthosEPS) for kindly providing particles for this research. WPB would like to thank SURF (www.surf.nl) for using the Dutch National Supercomputer Snellius for the DNS. This work is funded by the ERC Consolidator Grant No. 725183 “OpaqueFlows”.
\end{acknowledgments}

\appendix
\section{\label{A:1}Flow conditions of additional MRI experiments}
\begin{figure}[ht]
    \centering
    \includegraphics[width=0.5\textwidth]{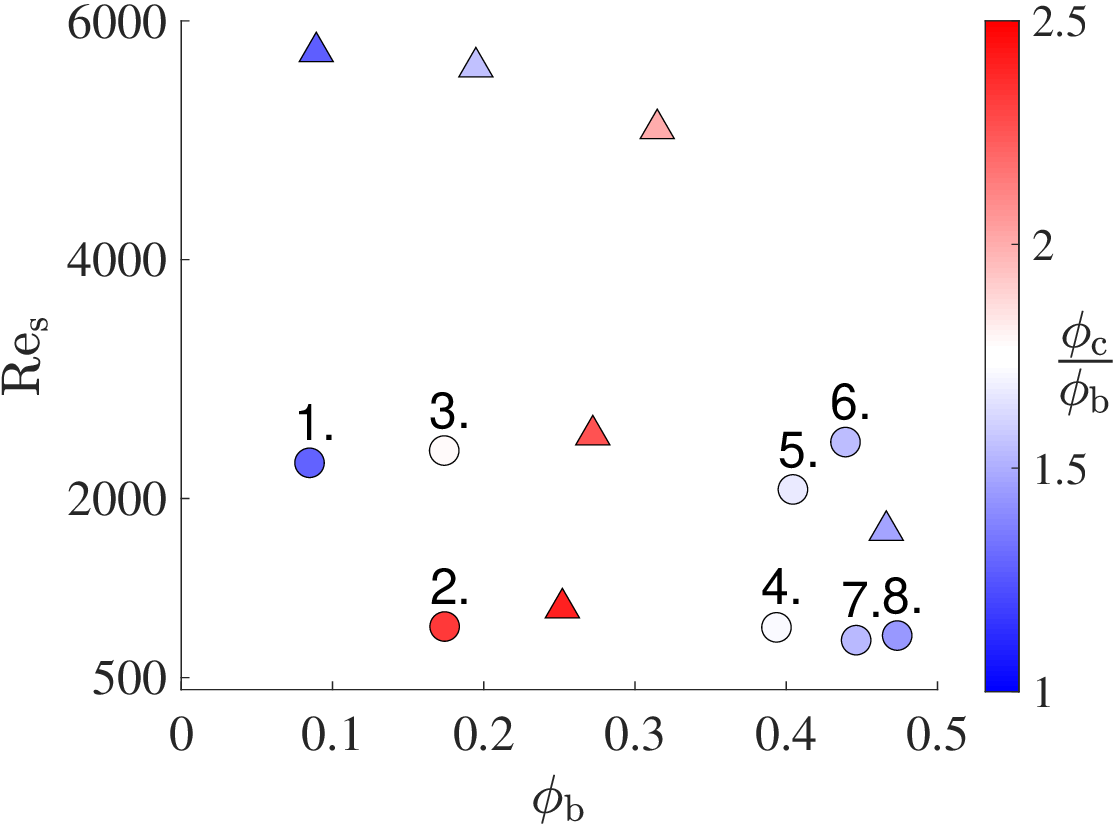}
    \caption{Similar regime map as shown in Fig.~\ref{fig:8a}, here the additional MRI experiments are numbered. The corresponding flow conditions are listed in Table \ref{table:a1}.}
    \label{fig:a1}
\end{figure}

This appendix list the flow conditions of the additional MRI experiments presented in Fig.~\ref{fig:8a}. The same MRI setup and protocols are used as described in Sec.~\ref{Sect:experimentaldetails}. The cases are listed for increasing $\phi_b$, as indicated in Fig.~\ref{fig:a1}. Here a number is added above the corresponding marker.

\begin{table}[htb]
	\centering
	\caption{Experimental conditions corresponding to the round markers in Fig.~\ref{fig:a1}.}
	\begin{tabular}[t]{p{1.8cm} p{1.5cm} p{1.4cm} p{2.1cm} p{1.5cm} p{1.5cm} p{1.5cm} p{1.5cm} p{2.7cm} p{1.4cm}}
	\toprule
	\textbf{Case}   &\textbf{$Re_l$}  &\textbf{$\phi_b$}  &\textbf{$\nu$ ($m^2/s$)} &\textbf{$u_{l,b} (m/s)$}  &\textbf{$\nu_s/\nu$} &\textbf{$Re_s$}    &\textbf{$u_{l,b}/u_{b}$} & \textbf{$f$}\\
	\midrule
	1	&  2895	    & 0.085	 & 1.43E--06	    & 0.1348	& 1.26	& 2299	 & 0.987	& 5.02E--2\\
    2	&  1566	    & 0.174	 & 1.43E--06	    & 0.0707	& 1.69	&  928	 & 0.960	& 7.92E--2\\
    3	&  4049	    & 0.174	 & 1.43E--06	    & 0.1844	& 1.69	& 2402	 & 0.966	& 5.05E--2\\
    4	&  4777	    & 0.394	 & 1.34E--06	    & 0.1938	& 5.19	&  921	 & 0.918	& 6.52E--2\\
    5	& 11694	    & 0.404	 & 1.42E--06	    & 0.5080	& 5.63	& 2076	 & 0.931	& 4.19E--2\\
    6	& 18692	    & 0.439	 & 1.38E--06	    & 0.8012	& 7.56	& 2473	 & 0.943	& 3.66E--2\\
    7	&  6574	    & 0.446	 & 1.29E--06	    & 0.2574	& 8.07	&  814	 & 0.923	& 6.55E--2\\
    8	&  9129	    & 0.473	 & 1.29E--06	    & 0.3601	& 10.7	&  852	 & 0.930	& 6.07E--2\\
	\bottomrule
    \end{tabular}
	\label{table:a1}
\end{table}

\bibliography{bibliography}

\end{document}